\documentclass{aa}  

\usepackage{graphicx}
\usepackage{txfonts}
\usepackage[switch]{lineno}
\begin{document}

\title{Extending the Event-weighted Pulsation Search to Very Faint Gamma-ray Sources}

\author{P. Bruel}

\institute{Laboratoire Leprince-Ringuet, Ecole polytechnique, CNRS/IN2P3, F-91128 Palaiseau, France\\
  \email{Philippe.Bruel@llr.in2p3.fr}
}

\date{Received 2 November 2018; accepted 17 December 2018}

\abstract
% context heading (optional), leave it empty if necessary  
{Because of the relatively broad angular resolution of current gamma-ray instruments in the MeV-GeV energy range, the photons of a given source are mixed with those coming from nearby sources or diffuse background. This source confusion seriously hampers the search for pulsation from faint sources.}
% aims heading (mandatory)
{Statistical tests for pulsation can be made significantly more sensitive when the probability that a photon comes from the pulsar is used as a weight. However, the computation of this probability requires knowing the spectral model of all sources in the region of interest, including the pulsar itself. This is not possible for very faint pulsars that are not detected as gamma-ray sources or whose spectrum is not measured precisely enough. Extending the event-weighted pulsation search to such very faint gamma-ray sources would allow improving our knowledge of the gamma-ray pulsar population.}
% methods heading (mandatory)
{We present two methods that overcome this limitation by scanning the spectral parameter space, while minimizing the number of trials. The first one approximates the source/background ratio yielding a simple estimate of the weight while the second one makes use of the full spatial and spectral information of the region of interest around the pulsar.}
% results heading (mandatory)
{We test these new methods on a sample of 144 gamma-ray pulsars already detected by the {\it Fermi} Large Area Telescope data. Both methods detect pulsation from all pulsars of the sample, including the ones for which no significant phase-averaged gamma-ray emission is detected.}
% conclusions heading (optional), leave it empty if necessary 
{}


\maketitle
%
%-------------------------------------------------------------------

\section{Introduction} \label{sec:intro}

One of the important contributions of the {\it Fermi} Large Area Telescope (LAT)~\citep{latinstrument} to gamma-ray astronomy is the detection of many pulsars. Prior to the {\it Fermi} launch in 2008, at most 10 gamma-ray pulsars were known~\citep{thompson2008}. Ten years later, more than 200 have been detected\footnote{\url{https://confluence.slac.stanford.edu/display/GLAMCOG/Public+List+of+LAT-Detected+Gamma-Ray+Pulsars}}, representing the most abundant Galactic source class at GeV energies~\citep{3FGL}. Detecting more pulsars, especially faint ones, is very important to characterizing as well as possible the gamma-ray pulsar population.

One way of finding gamma-ray pulsars is to search for periodic emission after phase-folding gamma-ray data with rotation ephemerides, obtained from contemporaneous timing observations conducted in radio or X-rays. Prior to {\it Fermi}'s launch, a Pulsar Timing Consortium (PTC) was organized to support LAT observations of pulsars by providing accurate pulsar ephemerides~\citep{PTC}. Another way of finding gamma-ray pulsars is to look at LAT catalog unassociated sources with a pulsar-like spectrum~\citep[see {\it e.g}][]{clarck2017,pleunis2017} but, by construction, it does not target very faint gamma-ray sources. We thus focus on the former approach in the rest of the paper.

The sensitivity to pulsation depends on the purity of the event sample, which depends on the level of background and on the instrument performance, especially its angular resolution. In the MeV-GeV energy range, the Galactic diffuse emission is bright and is the main source of background for candidate pulsars near the Galactic plane. Neighboring sources can also significantly increase the level of background because of the finite angular resolution of the instrument. The LAT Point Spread Function (PSF) at low energy is driven by multiple scattering of the electron and positron created by pair-conversion of the photon in the tracker. The 68\% containment angle ranges from 5.5~degrees at 100~MeV to 0.1~degrees at 30~GeV~\citep{latinstrument}. 

As shown in~\cite{BKR08} and~\cite{kerr2011}, it is possible to efficiently mitigate the background problem by weighting the pulsation statistical test with the probability of each event to come from the pulsar position. To compute the exact probability, one needs a complete description of the Region of Interest (RoI) around the pulsar (large enough to fully take into account the effects of the PSF), including the spectral model of all gamma-ray sources in the RoI. So one prerequisite is to be able to measure the spectrum of the candidate pulsar, which is usually done by performing a maximum-likelihood fit of the RoI. When the fit leads to a clear detection of the pulsar, its spectrum is well measured and the photon probabilities can be computed. On the contrary, a weak detection of the pulsar or obviously its non detection precludes a good measurement of its spectrum. As a consequence, current event-weighted methods to search for pulsation cannot be applied to very faint gamma-ray sources.

The only possible way to overcome this limitation is to scan over the pulsar spectral parameters in order to find the ones maximizing the statistical test used to search for periodic emission. The obvious drawback is that the resulting significance must be corrected for the number of trials performed during the scan. Consequently, the challenge of extending the event-weighted methods to very faint pulsars consists of minimizing the number of trials when scanning the spectral parameter space, while maximizing the pulsation sensitivity, {\it i.e.} getting as close as possible to the spectral parameters yielding the optimal weights.

After presenting in Section~\ref{sec:htestw} the statistical test we use for pulsation searches, we present in Section~\ref{sec:simple} a method using a simple weight definition that does not involve a detailed spatial nor spectral modelling of the RoI. In Section~\ref{sec:model}, we describe a more complex method that takes advantage of the complete knowledge of the RoI and the full capabilities of the instrument.

Both methods are tested on a sample of 144 known LAT pulsars. This sample comprises the 117 pulsars from the LAT Second Gamma-ray Pulsar Catalog (2PC)~\citep{2PC}, among which only two are not significant gamma-ray sources when analysing 8~years of LAT data. Adding 27 post-2PC detected pulsars \citep{hou2014,laffon2014,smith2017} to our sample increases the number of faint pulsars to 12. The results, based on the analysis of 8 year of Pass~8 SOURCE class data~\citep{pass8,pass8P8R3}, are presented and discussed in Section~\ref{sec:results}.

\section{The event-weighted H-test} \label{sec:htestw}

To test the signal periodicity, we use the weighted H-test derived by~\cite{kerr2011} from the original H-test proposed by~\cite{dejager1989}:
\begin{equation} \label{eq:htestw}
H_{mw} = \mathrm{max} \left[ Z_{iw}^2 -c \times (i-1)\right], \quad 1\leq i \leq m
\end{equation}
The weighted H-test definition involves the weighted $Z_{mw}^2$ test:
\begin{equation} \label{eq:Ztestw}
Z^2_{mw} = \frac{2}{\sum_{i=1}^{N} w_i^2} \times \sum_{k=1}^{m} (\alpha^2_{wk} + \beta^2_{wk})
\end{equation}
where
\begin{equation} \label{eq:walpha}
\alpha_{wk} = \sum_{i=1}^{N} w_i \cos{(2\pi k \phi_i)}
\end{equation}
and
\begin{equation} \label{eq:wbeta}
\beta_{wk} = \sum_{i=1}^{N} w_i \sin{(2\pi k \phi_i)},
\end{equation}
where the sums run over the list of $N$ photons with pulsar rotational phase $\phi_i$ and weight $w_i$.

In the following, we adopt the H-test parameter recommended values $m = 20$ and $c = 4$ (since they provide an omnibus test) and write $H_w$ instead of $H_{20w}$. We note that an obvious property of $Z^2_w$ and $H_w$ is that they are not sensitive to a global scaling of the photon weights.

The analytic, asymptotic null distribution of $H_w$ has been derived by~\cite{kerr2011}. Because we want to use $H_w$ even with small event samples, we derive this distribution directly with Monte Carlo simulations, allowing us to know, for any $x>0$, the probability $P(H_w>x)$, that $H_w$ evaluated on a non-periodic sample can fluctuate to values larger than $x$. The calibration procedure is described in Appendix~\ref{app:htestcalib} and Appendix~\ref{app:whtestcalib}. We show that $P(H_w>x)$ can be parameterized as a function of the sum of the weights, computed under the prescription that the maximum weight is 1.

Rather than using directly $H_w^\mathrm{meas}$, the measured value of $H_w$, when searching for pulsation, we use instead $P_w = -\log_{10}{P(H_w>H_w^\mathrm{meas})}$, that we name pulsation significance for simplicity's sake. We note that the relation between $P_w$ and $n_\sigma$, the significance expressed in number of $\sigma$, is given by $P_w = -\log_{10}\left(1-\mathrm{erf}(n_\sigma/\sqrt{2})\right)$. The 3, 4 and 5$\sigma$ levels correspond to $P_w \sim 2.57, 4.20$ and 6.24, respectively.

\section{Simple weights} \label{sec:simple}

Let us consider an RoI centered on a pulsar. The probability that a photon originates from the pulsar depends on the position and spectrum of all gamma-ray sources in the ROI and the response functions of the instrument. We assume that all sources are steady and that the pulsar spectrum does not depend on phase. For a photon at a position $\vec{\Omega}$ with an energy $E$, the probability that it comes from the pulsar is the ratio of the pulsar differential rate, $r_\mathrm{psr}$, and the total differential rate at this energy and position. Defining $r_\mathrm{bkg}$ as the sum of all but the pulsar differential rates, we have:
\begin{equation} \label{eq:wdefinition1}
  w(E,\vec{\Omega}) = \frac{r_\mathrm{psr} (E,\vec{\Omega})}{r_\mathrm{psr} (E,\vec{\Omega}) + r_\mathrm{bkg} (E,\vec{\Omega})}.
\end{equation}

\subsection{Simple weight definition} \label{sec:simple_weightdef}

Since we are interested in very faint pulsars, we assume that the pulsar differential rate is negligible compared to $r_\mathrm{bkg}$. Let us also assume that the background only comprises an isotropic diffuse emission. The background differential rate thus depends only on the energy $E$ and we have:
\begin{equation} \label{eq:wdefinition2}
  w(E,\vec{\Omega}) = r_\mathrm{psr} (E,\vec{\Omega})/r_\mathrm{bkg} (E).
\end{equation}

We note that the resulting weight $w$ is directly proportional to the pulsar absolute flux. Since the event-weighted statistical tests are unchanged when all weights are scaled by a constant, in the limit of very faint pulsars only the shape of the pulsar spectrum matters and not its absolute flux normalization. We also note that the weight position dependence only comes from the pulsar differential rate. Since the pulsar is a point source, the position dependence is described by the instrument PSF.

We can rewrite the weight as the product of two functions:
$$ w(E,\vec{\Omega}) = f(E) \times g(E,\vec{\Omega})$$
where $g(E,\vec{\Omega})$ contains all the position dependence and is defined such that it is 1 at the pulsar position at all energies. $f(E)$ is the weight at the pulsar position and depends on the pulsar and background spectral shapes and on the energy dependent part of the PSF normalization. The maximum of $f(E)$ is set to 1 to follow the maximum weight prescription mentioned in Section~\ref{sec:htestw}.

At a given energy, the LAT PSF is well approximated by a Moffat profile~\citep{moffat}\footnote{Formerly referred to as King profile in \url{https://fermi.gsfc.nasa.gov/ssc/data/analysis/documentation/Cicerone/Cicerone_LAT_IRFs/IRF_PSF.html}}:
\begin{equation} \label{eq:kingdef}
  K(x,s) = \frac{1}{4\pi s^2} \left( 1+\frac{x^2}{4 s^2} \right)^{-2}
\end{equation}
where $x$ is the angular distance to the source and $s$ a scale parameter.

The integral $\int_{0}^{\infty} \int_{0}^{2\pi} K(x,s) x \mathrm d x \mathrm d\theta = 1$ and we note that the integral up to a distance of $3s$ is about 0.68, so that the parameter $s$ corresponds to a third of the PSF 68\% containment angle. We can thus write the function $g(E,\vec{\Omega})$ as
\begin{equation} \label{eq:weightdef_g}
g(E,x) = \left( 1+\frac{9x^2}{4 \sigma^2_\mathrm{psf}(E)} \right)^{-2}
\end{equation}
where $\sigma_\mathrm{psf}(E)$ is the energy dependent PSF 68\% containment angle, which for the LAT Pass 8 SOURCE event class can be parameterized as follows~\citep{PSFDETERMINATION}:
$$\sigma_\mathrm{psf}(E) = p_0 (E/100)^{p_1} \oplus p_2$$
with $p_0 = 5.11$~degrees, $p_1 = -0.76$, $p_2 = 0.082$~degrees and $E$ in MeV and the addition is in quadrature.

The weight position-independent part, $f(E)$, corresponds to the weight at the position of the pulsar. Its computation is made simple by the fact that both the pulsar and background rates involve the same instrument effective area, which cancels out in Equation~\ref{eq:wdefinition2}. As a consequence $f(E)$ can be computed directly using the pulsar and background spectral shapes, as well as the energy dependent part of the PSF normalization ($1/4\pi s^2$ in Equation~\ref{eq:kingdef}). 

As shown in~\cite{2PC}, the spectral shape of pulsars between 60~MeV and 60~GeV energy range can be modeled with a power law with an exponential cutoff:
\begin{equation} \label{eq:PLEC}
  \mathrm dN/\mathrm dE \propto (E/E_0)^{-\gamma} e^{-(E/E_\mathrm{c})^\beta}
\end{equation}
with, for most of pulsars, an index $\gamma$ ranging from 0.5 to 2, an energy cutoff $E_\mathrm{c}$ between 0.6 and 6~GeV (and $\beta$ set to 1 in 2PC). We consider three $(\gamma,E_\mathrm{c})$ test-cases: (2.0, 0.6~GeV), (0.5, 0.6~GeV) and (2.0, 6~GeV). The last two encompass the bulk of the pulsars, while the first one corresponds to pulsars whose pulsation is driven by low energy photons.

Regarding the background spectral shape, we use the spectrum of the Galactic diffuse emission towards the Galactic center. It can be modeled with a smoothly broken power law, with spectral indices of $\sim 1.6$ and $\sim 2.5$ below and above $\sim 3$~GeV, respectively\footnote{The spectral indices are estimated using the {\it Fermi}-LAT Galactic diffuse emission model {\tt gll\_iem\_v06.fits} available at \url{https://fermi.gsfc.nasa.gov/ssc/data/access/lat/BackgroundModels.html}.}.

\begin{figure}[ht]
  \centering
  \includegraphics[width=9.5cm]{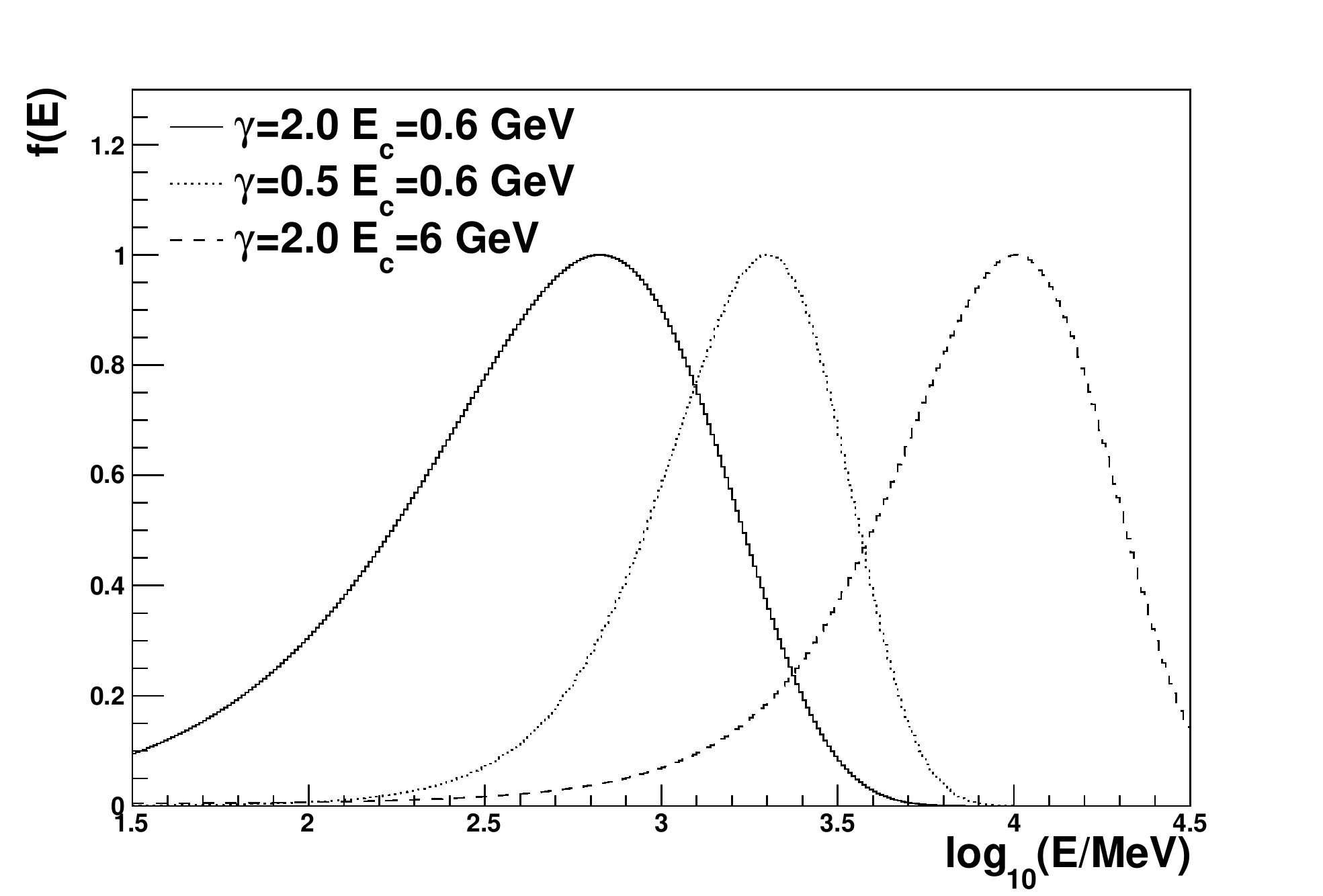}
  \caption{The weight position independent part $f(E)$ for three test pulsar spectra.}
  \label{fig:simpleweight_fE_testcases}
\end{figure}

Figure~\ref{fig:simpleweight_fE_testcases} shows $f(E)$ for the three pulsar test-cases. Expressed as a function of $\log_{10}E$, the function $f$ is gaussian-like. The pulsar energy cutoff is responsible for the decrease of $f$ at high energy. At low energy, its decrease is mainly driven by the increase of the instrument PSF. The center position parameter $\mu_{w}$ is located between 2.5 and 4.5 (corresponding to 30~MeV and 30~GeV, respectively) while the half 68\% width parameter $\sigma_{w}$ ranges from 0.3 to 0.45.

This result does not change qualitatively when using the spectrum of the Galactic diffuse emission at high Galactic latitude: for a given pulsar spectrum, changing the background spectral index moves the peak position of $f(E)$ but its shape remains gaussian-like, with a width in the same [0.3,4.5] range. As a consequence, we can use the same weight definition for young pulsars, that are mostly near the Galactic ridge, and millisecond pulsars, that tend to lie at high Galactic latitude.

We approximate $f(E)$ with a Gaussian in $\log_{10}E$:
\begin{equation} \label{eq:fEdef}
 f(E) = e^{-0.5(\log_{10}E-\mu_w)^2/\sigma_{w}^2}.
\end{equation}

We note that the examples in Figure~\ref{fig:simpleweight_fE_testcases} are not exactly gaussian. Using other functional forms that better model the negative tail of $f(E)$ does not improve the performance of the pulsation search, that is, better matching of $f(E)$ does not compensate for the simple assumptions that we use (very faint source on top of an isotropic background). Similarly, in Section~\ref{sec:psfevttype} we show that refining the PSF information fails to improve the performance.

The gaussian definition of $f(E)$ introduces two parameters, $\mu_{w}$ and $\sigma_{w}$. Formally, one would have to scan over these two parameters in order to fully explore the pulsar spectral parameter space. However, $\sigma_{w}$ varies much less than $\mu_{w}$, allowing us to fix $\sigma_{w}$ and to vary only $\mu_{w}$. We note that the $\sigma_{w}$ range was obtained under the very faint pulsar hypothesis. For brighter pulsars, the energy range over which the pulsar emission is significant is larger, leading to a wider shape for $f(E)$, which corresponds to larger values of $\sigma_{w}$. For this reason, we choose to fix $\sigma_{w}$ to 0.5. We check that, on average, this choice is valid for all pulsars in Section~\ref{sec:results}.

Thus, the simple weight definition that we propose is:
\begin{equation} \label{eq:weightdef}
 w(E,x,\mu_w) = e^{-2(\log_{10}E-\mu_w)^2} \left( 1+\frac{9x^2}{4 \sigma^2_\mathrm{psf}(E)} \right)^{-2}
\end{equation}

\subsection{Simple weight scan}

\begin{figure}[ht]
  \centering
  \includegraphics[width=9.5cm]{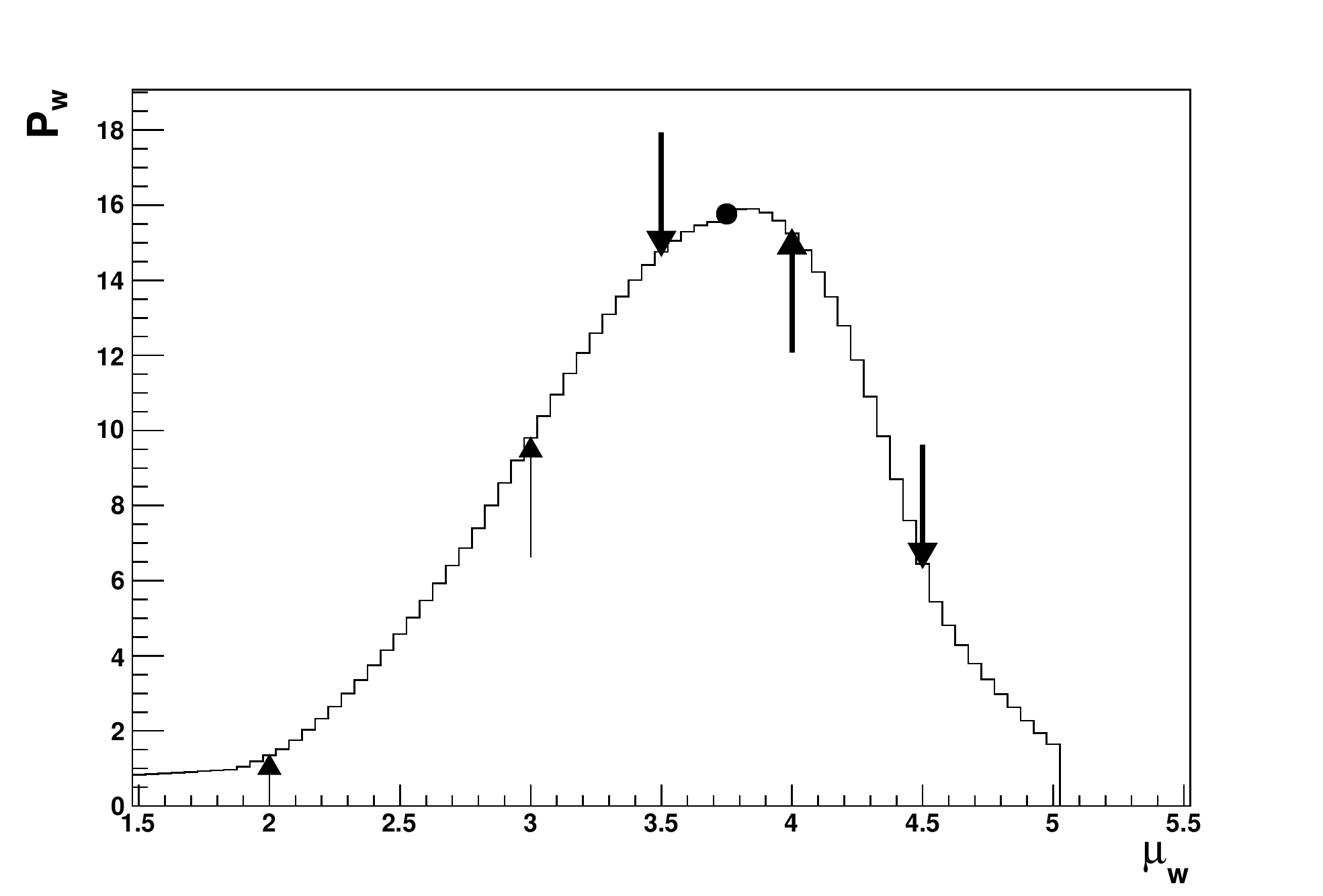}
  \caption{The pulsation significance as a function of $\mu_{w}$ for PSR~J1646$-$4346. The upward arrows correspond to the first three trials, while the downward arrows correspond to the two subsequent trials. The sixth trial, shown with a circle, is the maximum of the gaussian passing through the three points corresponding to the thick-line arrows.}
  \label{fig:simpleweight_algo}
\end{figure}

For a given value of $\mu_w$, we can compute $H_w$ and the corresponding pulsation significance $P_w$ using the weights $w(E,x,\mu_w)$. Figure~\ref{fig:simpleweight_algo} shows $P_w$ as a function of $\mu_w$ for PSR~J1646$-$4346~\citep{smith2017}. We chose this pulsar to illustrate the method because it is one of the pulsars in our sample that are not detected as a gamma-ray source with 8~years of LAT data and, as a consequence, for which the original~\cite{kerr2011} method can not be applied.

\begin{figure}[ht]
  \centering
  \includegraphics[width=9.5cm]{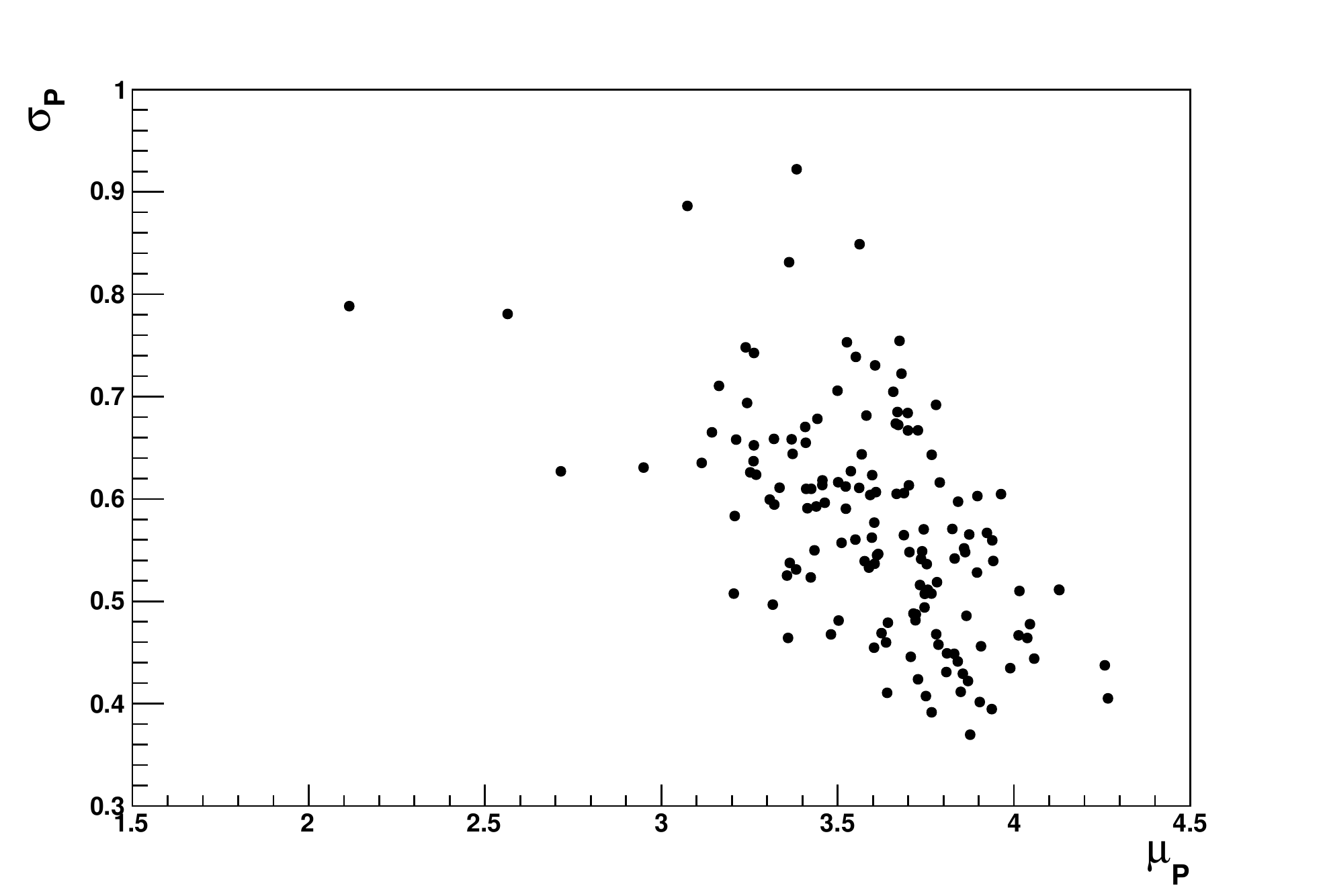}
  \caption{The 68\% width as a function of the peak position of $P_w(\mu_w)$ for the 144 pulsar sample.}
  \label{fig:simpleweight_PPdist_2PC}
\end{figure}

As can be seen in this example, $P_w(\mu_w)$ has a gaussian shape around its maximum. This shape is simply the result of the relative matching between the weights $w(E,x,\mu_w)$ and the optimal weights, when $\mu_w$ varies. We name $\mu_{P}$ and $\sigma_{P}$ the maximum position and the 68\% width of $P_w(\mu_w)$. Figure~\ref{fig:simpleweight_PPdist_2PC} shows the distribution of $\sigma_{P}$ versus $\mu_{P}$ for the 144 pulsar sample. The bulk of the pulsars have $3<\mu_{P}<4$ while a few have a low $\mu_{P}$, corresponding to a pulsation signal driven by low energy photons ({\it e.g.} PSR~J1513$-$5908, aka PSR~B1509$-$58~\citep{B1509,kuiper2017}, whose energy cutoff is below 30~MeV). The width $\sigma_{P}$ is between 0.35 and 0.9.

Searching for pulsation consists in finding the maximum of $P_w(\mu_w)$. The easiest way to do so is to perform a very fine scan over $\mu_w$ but that would lead to a too large number of trials. Taking advantage of the gaussian-shape of $P_w(\mu_w)$ around its maximum, we perform the following algorithm (illustrated in Figure~\ref{fig:simpleweight_algo}):
\begin{enumerate}
\item Test three values of $\mu_w = (2,3,4)$. Let $\mu_0$ be the one giving the maximum $P_w$.
\item Test two more values of $\mu_w = (\mu_0-0.5,\mu_0+0.5)$. The 0.5 distance is chosen such that it is of the order of the minimum of the 68\% width of $P_w(\mu_w)$. Let $\mu_1$ be the one giving the maximum of $P_w$ among all the trials with $2\leq \mu \leq 4$;
\item Test $\mu_w = \mu_g$, the position of the maximum of the gaussian passing through the three points $\mu_1-0.5,\mu_1,\mu_1+0.5$ (all tested in the previous steps);
\end{enumerate}

The resulting pulsation significance of the simple method, corrected for the 6 trials, is:
\begin{equation} \label{eq:Psdef}
P_{s} = \max_{\mu_{w} \; \mathrm{scan}} P_w - \log_{10}6
\end{equation}

We note that the 6 trials are partially correlated, especially the ones with a small difference in $\mu_{w}$. Unfortunately there is no simple way to estimate and take into account these correlations when correcting the pulsation significance and so we simply ignore them and adopt the conservative definition of $P_{s}$ given by Equation~\ref{eq:Psdef}.

\section{Model weights} \label{sec:model}

To increase the sensitivity of the pulsation search, one has to go beyond the simple weight definition of Equation~\ref{eq:weightdef}. There are two possible approaches:
\begin{itemize}
\item use the spatial and spectral description of all sources in the RoI;
\item take into account the full PSF information.
\end{itemize}

While the former requires precise RoI modelling including spectral fits, the latter can be achieved by a minimal change of the weight definition, as it is a consequence of a refinement of the PSF definition.

\subsection{PSF event types} \label{sec:psfevttype}

One of the new features of Pass~8~\citep{pass8}, the latest version of the reconstruction and selection of LAT data, is the introduction of the PSF event types. Prior to Pass~8, the events of a given class ({\it e.g.} SOURCE class) were divided into two subclasses, FRONT and BACK, depending on the location of the photon pair-conversion in the LAT tracker. Because of the different thicknesses of the converters in these two sections of the tracker, the PSF of the two subclasses are significantly different: at 1~GeV the FRONT and BACK PSF 68\% containment angles are 1.2 and 0.61~degrees, respectively. In Pass~8, thanks to a multivariate analysis, the events are divided into four event types PSF0, PSF1, PSF2 and PSF3, ordered in decreasing quality of the reconstructed direction, with a PSF 68\% containment angle at 1~GeV of 1.8, 1.0, 0.66 and 0.42~degrees, respectively.

It is straightforward to take into account the PSF events types in the context of the simple weight definition of Equation~\ref{eq:weightdef}: one has just to properly modify the $f(E)$ part of the weight induced by the $1/4\pi s^2$ factor of the normalization of the Moffat profile (Equation~\ref{eq:kingdef}). We note that this modification is made simple by both the faint-source and uniform-background hypotheses.

Tested on LAT data, this modification failed to clearly increase the pulsation significance of known pulsars. This failure can be explained by two reasons. First, as soon as the faint-source hypothesis does not hold, the correction induced by the $1/4\pi s^2$ factor of the Moffat profile normalization is not correct: in the extreme situation in which the background is negligible, all photons should have a weight $\sim 1$, regardless of their PSF event type.

Moreover, taking into account the better PSF description given by the PSF event types is in fact equivalent to a better spatial description of the RoI. This improvement might not be large enough to compensate the crude approximation of the uniform-background hypothesis. In other words, one has to use a precise spatial and spectral description of the RoI to fully take advantage of the additional information provided by the PSF event types.

As a consequence, going beyond the simple weights presented in the previous section requires the full spatial and spectral description of the RoI around the pulsar.

\subsection{Pulsar RoI description} \label{sec:spectralfit}

To compute the weights for each pulsar, we perform a binned maximum-likelihood fit, following the standard {\it Fermi}-LAT analysis procedure, of the $10.05\times10.05$~degrees RoI centered on the pulsar, with a pixel size of 0.05~degrees. We use 37 bins in $\log_{10}E$, between 63.1~MeV and 316.2~GeV.

We include in the source model the following components:
\begin{itemize}
\item the Galactic diffuse emission and the isotropic template (that accounts for the isotropic diffuse emission as well as the residual background)~\footnote{\url{https://fermi.gsfc.nasa.gov/ssc/data/access/lat/BackgroundModels.html}};
\item all point-like and extended sources from the preliminary 8-year {\it Fermi}-LAT source list (FL8Y)~\footnote{\url{https://fermi.gsfc.nasa.gov/ssc/data/access/lat/fl8y/}}, within 5~degrees of the RoI border.
\end{itemize}
If the closest point source to the pulsar is within a distance of 1.5 times the source 95\% error radius, we consider the source to be the pulsar and set its position to the pulsar position. Otherwise we add a new source at the pulsar position.

We perform the maximum-likelihood fit with SOURCE class events (combining all PSF event types) above 160~MeV whose zenith angle is less than 90~degrees to avoid Earth limb contamination, collected between 2008 August~4 and 2016 August~2. The significance of each source in the model is estimated using the Test Statistic, TS, defined as twice the difference in log-likelihood obtained with and without the source. A $\mathrm{TS}=25$ corresponds to $\sim 4\sigma$ significance~\citep{mattox}. The 160~MeV energy threshold for the maximum-likelihood fit has been chosen to mitigate the effect of systematic errors due to our imperfect knowledge of the Galactic diffuse emission.
%We note that, in the FL8Y analysis, this issue was taken care of with the introduction of weights in the definition of the likelihood.

Using the model derived by the fit, we build for each PSF event type the 3-D map (sky position and energy) of the number of predicted events coming from the pulsar as well as the 3-D map of the total number of predicted events. We are then able to compute the weights by simply dividing the 2 maps.

When computing $H_w$, we assign to each event the weight of the bin corresponding to the event position and energy of the 3-D map corresponding to the PSF event type of the event.

In~\cite{kerr2011}, the weights are computed using the {\it Fermi}-LAT Science Tool {\tt gtsrcprob}\footnote{\url{https://fermi.gsfc.nasa.gov/ssc/data/analysis/scitools/overview.html}} that assigns to each event the probability that the event belongs to a given source of the RoI. Our binned approach is less CPU intensive and the chosen binning is fine enough to not induce any significant loss of sensitivity.

\subsection{Spectral parameter scan}

For bright gamma-ray pulsars, the maximum-likelihood fit gives a large TS for the pulsar and its spectral parameters are well estimated. We can thus use these spectral parameters to compute the weights, as done in~\cite{kerr2011}. On the contrary, for faint pulsars (TS<25) or pulsars just above the TS=25 threshold for which the spectrum is not precisely estimated, we have to scan over the spectral parameters.

Instead of using Equation~\ref{eq:PLEC} to model the power law with an exponential cutoff, we use the following expression:
\begin{equation} \label{eq:PLEC2}
  \mathrm dN/\mathrm dE \propto (E/E_0)^{-\gamma} e^{a(E_0^\beta-E^\beta)}
\end{equation}
as used in FL8Y and the forthcoming 4FGL catalog, with $\beta$ given by the fit for bright pulsars or fixed to $2/3$ for faint or not-detected ones. The formal change between the energy cut off and parameter $a$ is convenient to scan the spectral parameter space, as will be shown later.

\begin{figure}[ht]
  \centering
  \includegraphics[width=9.5cm]{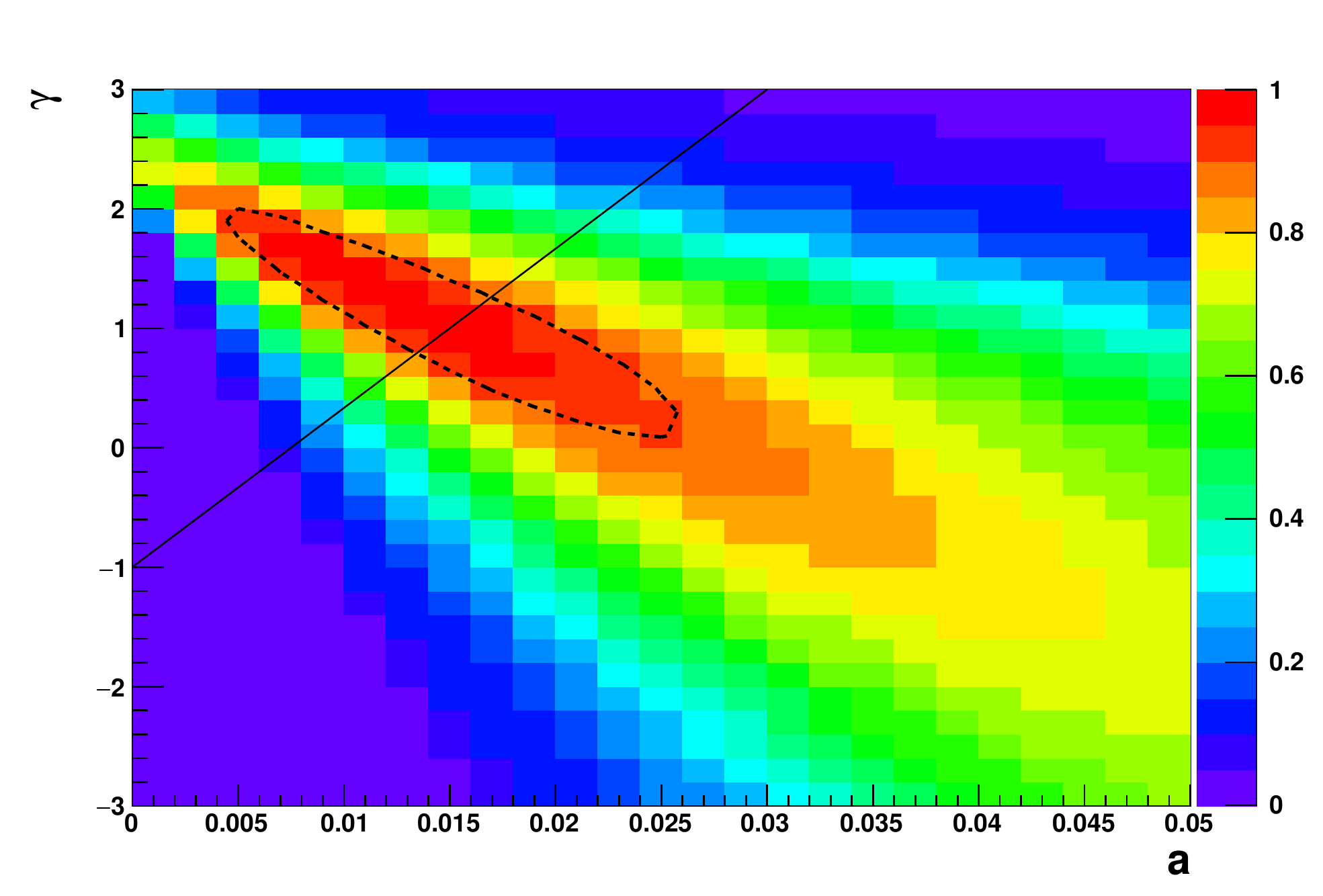}
  \caption{Pulsation significance in the $(a,\gamma)$ parameter space normalized relative to its maximum for PSR~J1646$-$4346. The dashed contour line corresponds to the 90\% level. The solid straight line corresponds to the line $L_m$ along which a scan is performed to obtain the model weight pulsation significance, as described in Section~\ref{sec:modelweightscan}.}
  \label{fig:model_scan2d}
\end{figure}

For a given set of $(a,\gamma)$, we fix the corresponding spectral parameters and perform the maximum-likelihood fit with the pulsar normalization being the only free parameter. If $\mathrm{TS}<4$, we set the normalization to the 68\% confidence limit. Using the resulting pulsar spectrum, we compute the four PSF event type weight maps and compute the pulsation significance, taking into account the PSF event type information by using for each event the PSF event type weight map corresponding to the event. Figure~\ref{fig:model_scan2d} shows how the pulsation significance varies in the $(a,\gamma)$ plane for PSR~J1646$-$4346.

\begin{figure}[ht]
  \centering
  \includegraphics[width=9.5cm]{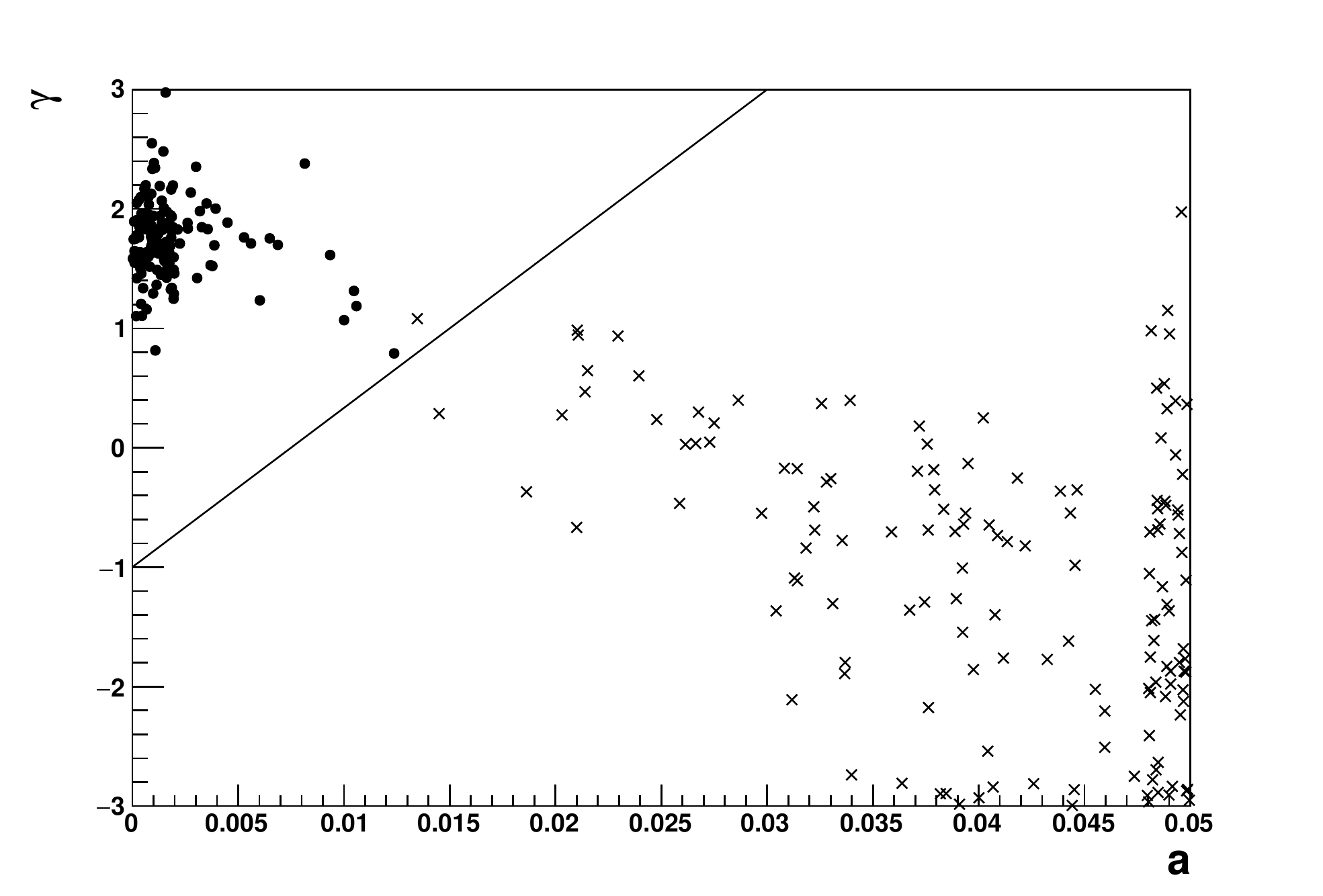}
  \caption{Position of the lowest (dots) and highest (crosses) $a$ points of ${\cal E}_{90}$ (the region of the $(a,\gamma)$ plane with $P_w$ larger than 90\% of the maximum) for the 144 pulsar sample. The solid line corresponds to the line $L_m$ along which a scan is performed to obtain the model weight pulsation significance, as described in Section~\ref{sec:modelweightscan}.}
  \label{fig:model_lowerupper90}
\end{figure}

The iso-level contours have an ellipse-shape. Let us consider the high $P_w$ region corresponding to $P_w$ larger than 90\% of the maximum, whose ellipse-like contour is named ${\cal E}_{90}$. To characterize ${\cal E}_{90}$, we use the lowest and highest $a$ extremities of ${\cal E}_{90}$. These points are shown in Figure~\ref{fig:model_lowerupper90} for the 144 pulsar sample. The important result is that the lowest and highest $a$ points of ${\cal E}_{90}$ are not mixed and it is possible to define a line separating them.

The fact that ${\cal E}_{90}$ is clearly elongated along a direction that tends to go from the upper-left to the lower-right of the $(a,\gamma)$ plane is the consequence of how the weights vary with the spectral parameters. For a given background spectrum, when $a$ increases, the effective position of the spectrum cutoff decreases and the energy position of the maximum weight at the pulsar position shifts to lower energy. On the other hand, the maximum weight energy position shifts to higher energy when $\gamma$ decreases. So when $a$ increases and $\gamma$ decreases, the two effects partly counterbalance each other and the maximum weight energy position does not vary very much. The situation is reversed along the minor axis of ${\cal E}_{90}$: the increases of $a$ and $\gamma$ both shift the maximum weight energy position to lower energy. This is why ${\cal E}_{90}$ is very eccentric.

We note that the convenient elliptical shape of the high $P_w$ region is the result of the change from Equation~\ref{eq:PLEC} to Equation~\ref{eq:PLEC2} to model the power law with an exponential cutoff.

\subsection{Model weight scan}  \label{sec:modelweightscan}

As for the simple weight method, the goal is to find the maximum of $P_w$ in the lowest possible number of trials. The characterization of the high $P_w$ region obtained in the previous section naturally suggests a 2-step procedure: first finding the major-axis of ${\cal E}_{90}$ then scanning along the major-axis to find the maximum $P_w$.

To find the major-axis of ${\cal E}_{90}$ we perform a scan along the line $L_m$, going from ($a=0,\gamma=-1$) to ($a=0.03,\gamma=3$), that crosses almost all ${\cal E}_{90}$, as shown in Figure~\ref{fig:model_lowerupper90}. The choice of $L_m$ is motivated by the fact that increasing both $a$ and $\gamma$ at the same time allows a fast variation of the maximum weight energy position and, therefore, an efficient exploration of the pulsar spectrum parameter space. This choice might be slightly refined in the future thanks to the analysis of a larger sample of gamma-ray pulsars.

\begin{figure}[ht]
  \centering
  \includegraphics[width=9.5cm]{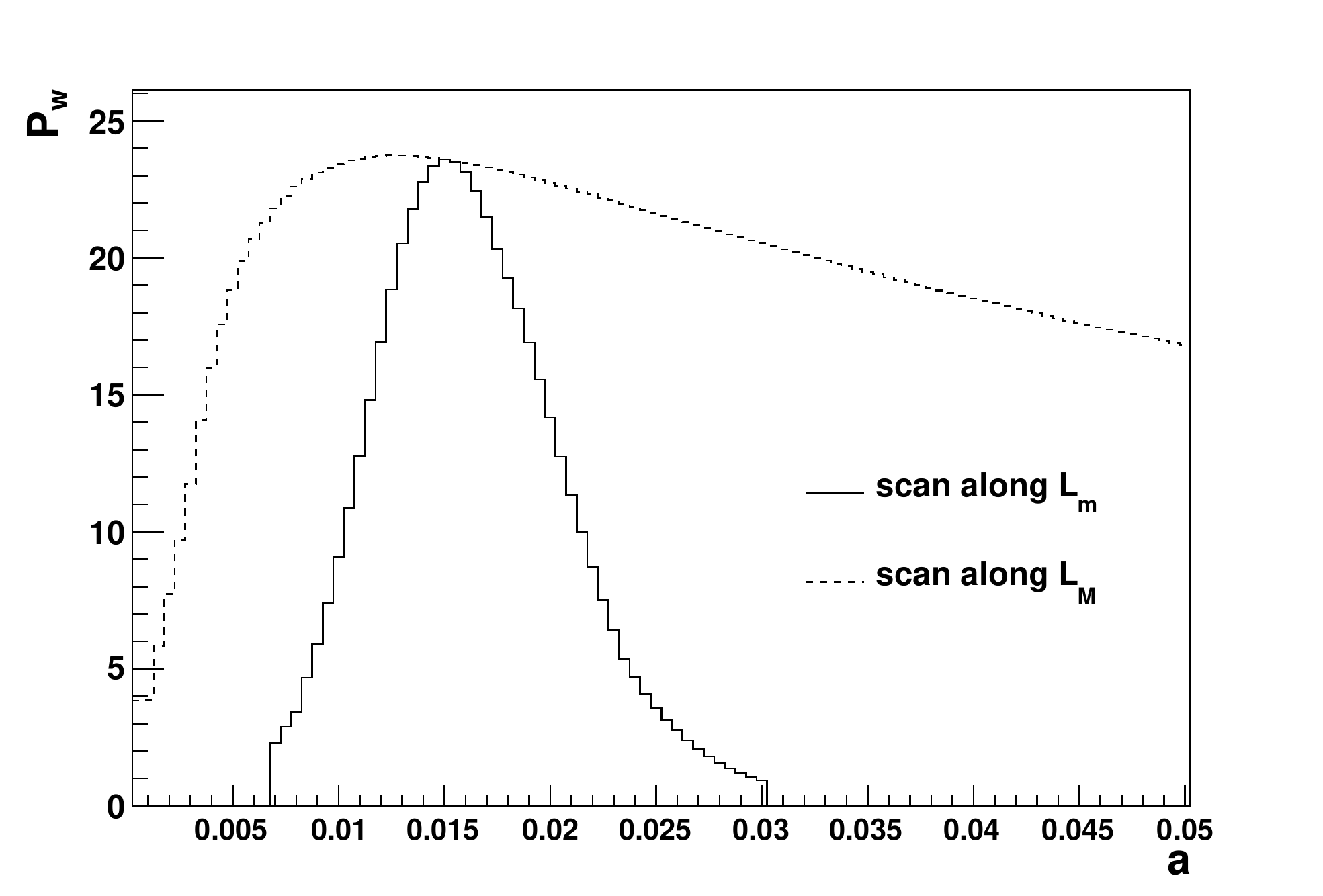}
  \caption{Pulsation significance as a function of $a$, when scanning along $L_m$ (solid line) and along $L_M$ (dashed line) for PSR~J1646$-$4346. $L_m$ and $L_M$ are two straight line in the pulsar parameter space $(a,\gamma)$. $L_m$ goes from ($a=0,\gamma=-1$) to ($a=0.03,\gamma=3$) while $L_M$ approximates the ${\cal E}_{90}$ major-axis.}
  \label{fig:model_step1step2}
\end{figure}

An example of the $L_m$ scan is shown in Figure~\ref{fig:model_step1step2}. Because the $P_w$ variation along $L_m$ is gaussian-like around the maximum, we can perform a similar 6-trial algorithm as for the simple weight method, with the following test positions:
\begin{enumerate}
\item Test three values of $a = (0.005,0.015,0.025)$. Let $a_0$ be the one giving the maximum $P_w$.
\item Test two more values of $a = (a_0-0.005,a_0+0.005)$. The 0.005 distance is chosen such that it is of the order of the minimum of the 68\% width of $P_w(a)$ when scanning along $L_m$. Let $a_1$ be the one giving the maximum of $P_w$ among all the trials with $0.005 \leq a \leq 0.025$;
\item Test $a = a_g$ the position of the maximum of the gaussian passing through the three points $a_1-0.005,a_1,a_1+0.005$  (all tested in the previous steps);
\item Let $A_m(a_m,\gamma_m)$ be the point giving the maximum $P_{w}$ among the 6 trials.
\end{enumerate}

\begin{figure}[ht]
  \centering
  \includegraphics[width=9.5cm]{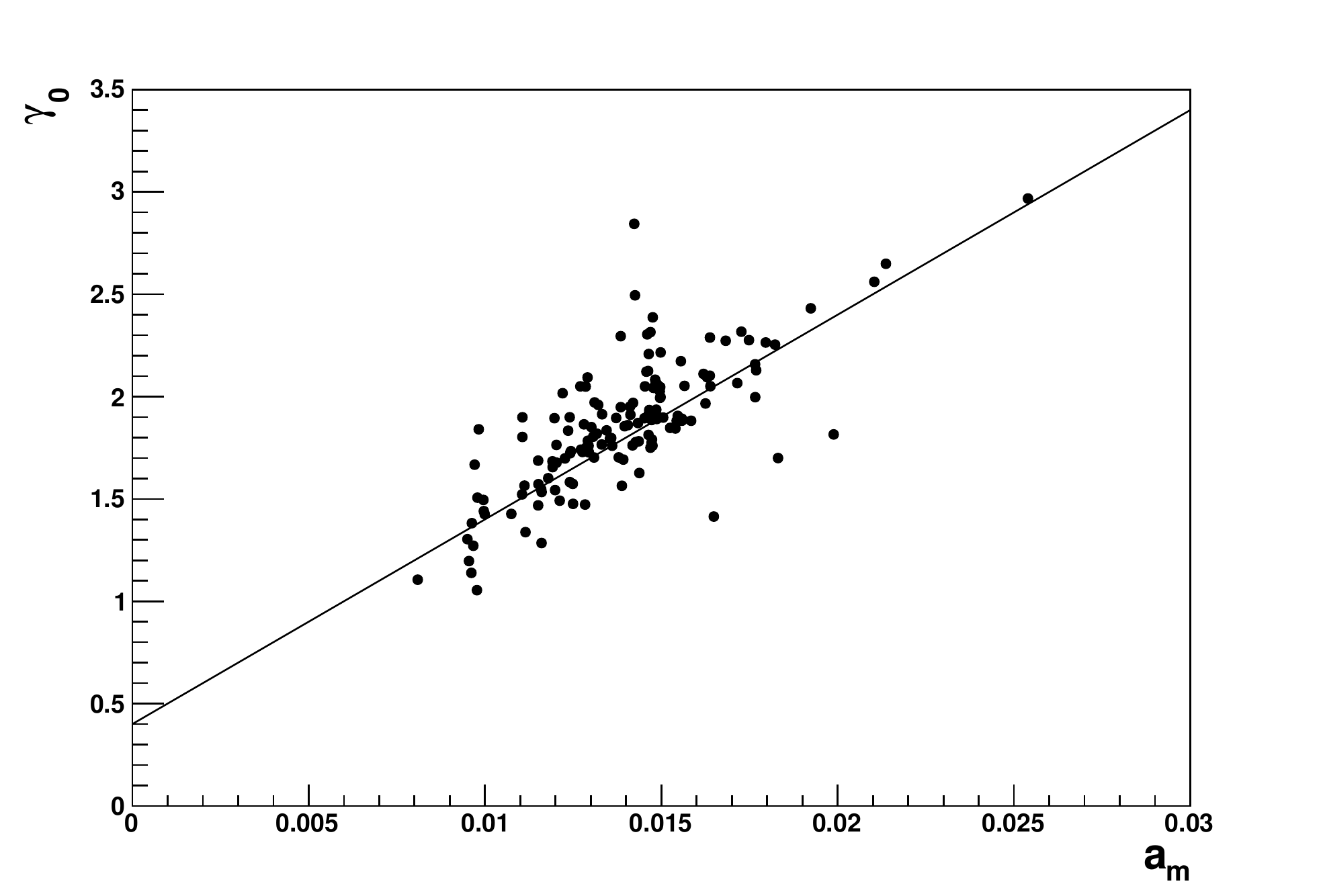}
  \caption{Correlation between $\gamma_0$, the intercept of the ${\cal E}_{90}$ major axis with the $a=0$ axis, and $a_m$, the position of the maximum $P_{w}$ along $L_m$ for the 144 pulsar sample. The solid line shows the $\gamma_0 = 0.4+100 a_m$ parameterization.}
  \label{fig:model_majoraxisintercept}
\end{figure}

$A_m$ is expected to lie on the major-axis and we need one more point to define it. Figure~\ref{fig:model_majoraxisintercept} shows $\gamma_0$, the intercept of the major-axis with the $\gamma$-axis, as a function of $a_m$. The correlation between the two can be modeled with $\gamma_0 = 0.4+100 a_m$. We use this correlation to choose the point $A_0(a,\gamma) = (0,0.4+100 a_m)$ that, together with $A_m$, defines the major axis $L_M$.

\begin{figure}[ht]
  \centering
  \includegraphics[width=9.5cm]{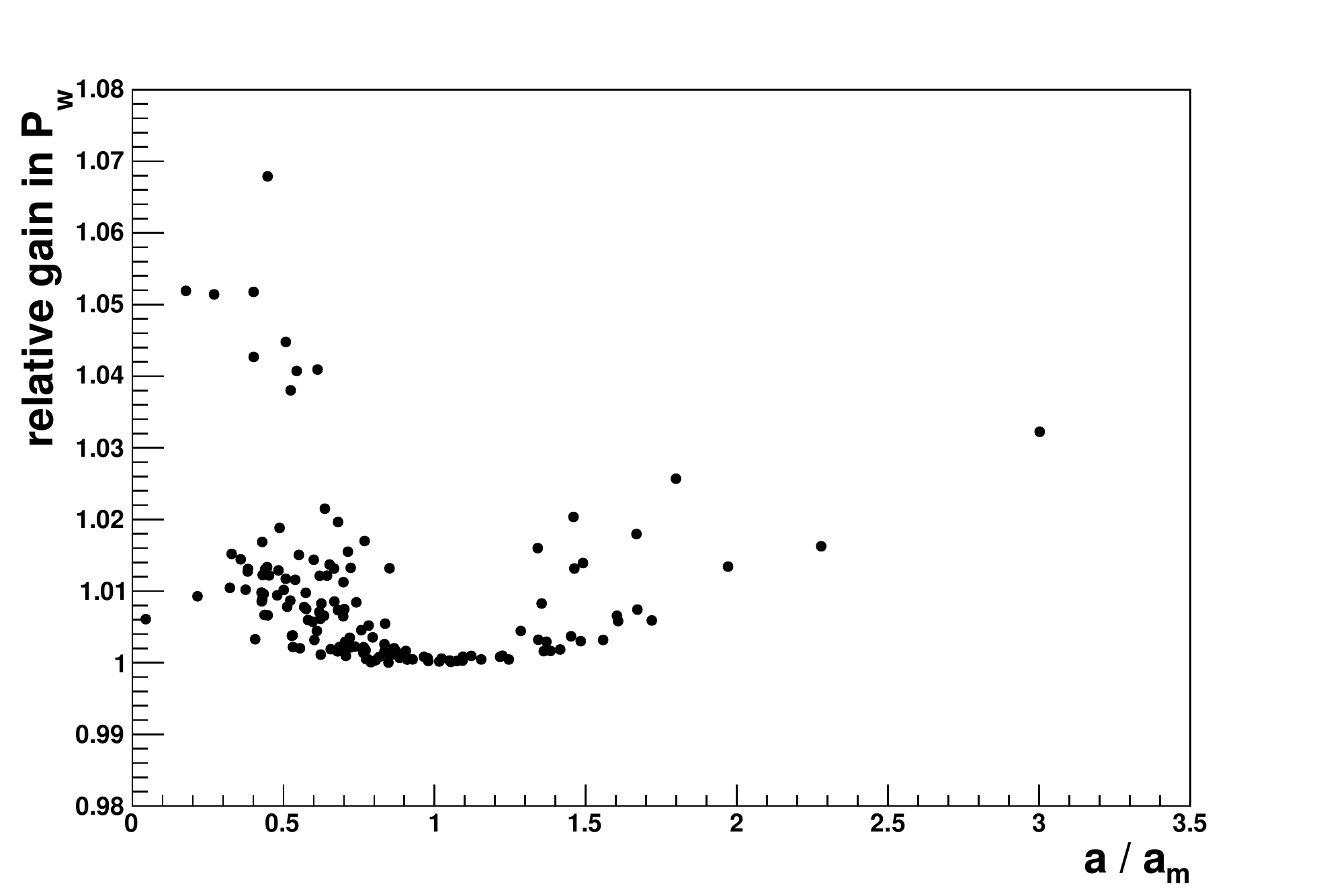}
  \caption{Maximum $P_w$ gain along $L_M$ as a function of its relative position with respect to $a_m$ for the 144 pulsar sample.}
  \label{fig:model_optstep2}
\end{figure}

An example of the $P_w$ variation along $L_M$ is shown in Figure~\ref{fig:model_step1step2}. The $L_M$ scan allows us to reach a higher $P_w$ than the $L_m$ scan, but, because the latter goes through ${\cal E}_{90}$, the gain in $P_w$ is rather modest (less than 10\% by definition of ${\cal E}_{90}$). To optimize the definition of the second step, we look at the maximum relative gain in $P_w$ along $L_M$  versus its relative position with respect to $a_m$. As can be seen in Figure~\ref{fig:model_optstep2}, the gain in $P_w$ is on average about 1\%, reaching at most 5\% for few pulsars. Regarding the optimal position $a$, it can be either smaller or larger than $a_m$. This would imply at least two more trials, on top of the six ones already performed during the $L_m$ scan, inducing an additional trial correction of $\log_{10}(8/6) = 0.125$.

For pulsars on the verge of pulsation detection, {\it i.e.} at the $4\sigma$ level, corresponding to $P_w \sim 4.2$, the additional trial correction corresponds to about 2\% of $P_w$, larger than the average 1\% gain of the $L_M$ scan. As a consequence, we choose not to perform the $L_M$ scan. The model weight pulsation significance corrected for the 6 trials is thus:
\begin{equation}
P_{m} = \max_{L_m \; \mathrm{scan}} P_w - \log_{10}6
\end{equation}

As in the simple weight method, the 6 trials are partially correlated and, for the same reasons, we choose to ignore them and to use a conservative definition of $P_{m}$.

\section{Results} \label{sec:results}

To test the performance of the simple and model weight methods, we apply them to the sample of 144 LAT pulsars (117 pulsars from 2PC and 27 post-2PC detected pulsars). For each pulsar, we perform the binned maximum-likelihood fit presented in Section~\ref{sec:spectralfit} to estimate the TS of the pulsar and compute the model weights. 12 pulsars are found to have $\mathrm{TS}<25$.

To perform the pulsation search, we select Pass 8 SOURCE class events within 5~degrees of the pulsar above 60~MeV. We use the {\it Fermi} plugin~\citep{ray2006} to the pulsar timing software \textsc{Tempo2}~\citep{hobbs2006,edwards2006} and the pulsar ephemeris provided by the PTC~\citep{PTC,smith2018} to convert the event arrival time into a pulsar rotational phase. The pulsation probabilities are computed using only the data collected during the validity period of the ephemerides.

We also compare the simple and model weight results to the one obtained with the original~\cite{kerr2011} method, which simply corresponds to the model weight method with the weights computed using the spectral parameters given by the maximum-likelihood fit. This is possible only when the source is significantly detected by the fit, {\it i.e.} when $\mathrm{TS}>25$. We name $P_\mathrm{fit}$ the corresponding pulsation significance. Compared to $P_m$, $P_\mathrm{fit}$ has the obvious advantage of being the result of only one trial.

To compare the performance of the new methods to a standard unweighted pulsation search, we also perform a grid search, with each grid point corresponding to a set of cuts in energy and distance to the pulsar. We use the grid parameters defined by~\cite{kerr2011}, where it is reported that the event weighting method improves the pulsation sensitivity by a factor 1.5--2.

\subsection{Model weights}

\begin{figure}[ht]
  \centering
  \includegraphics[width=9.5cm]{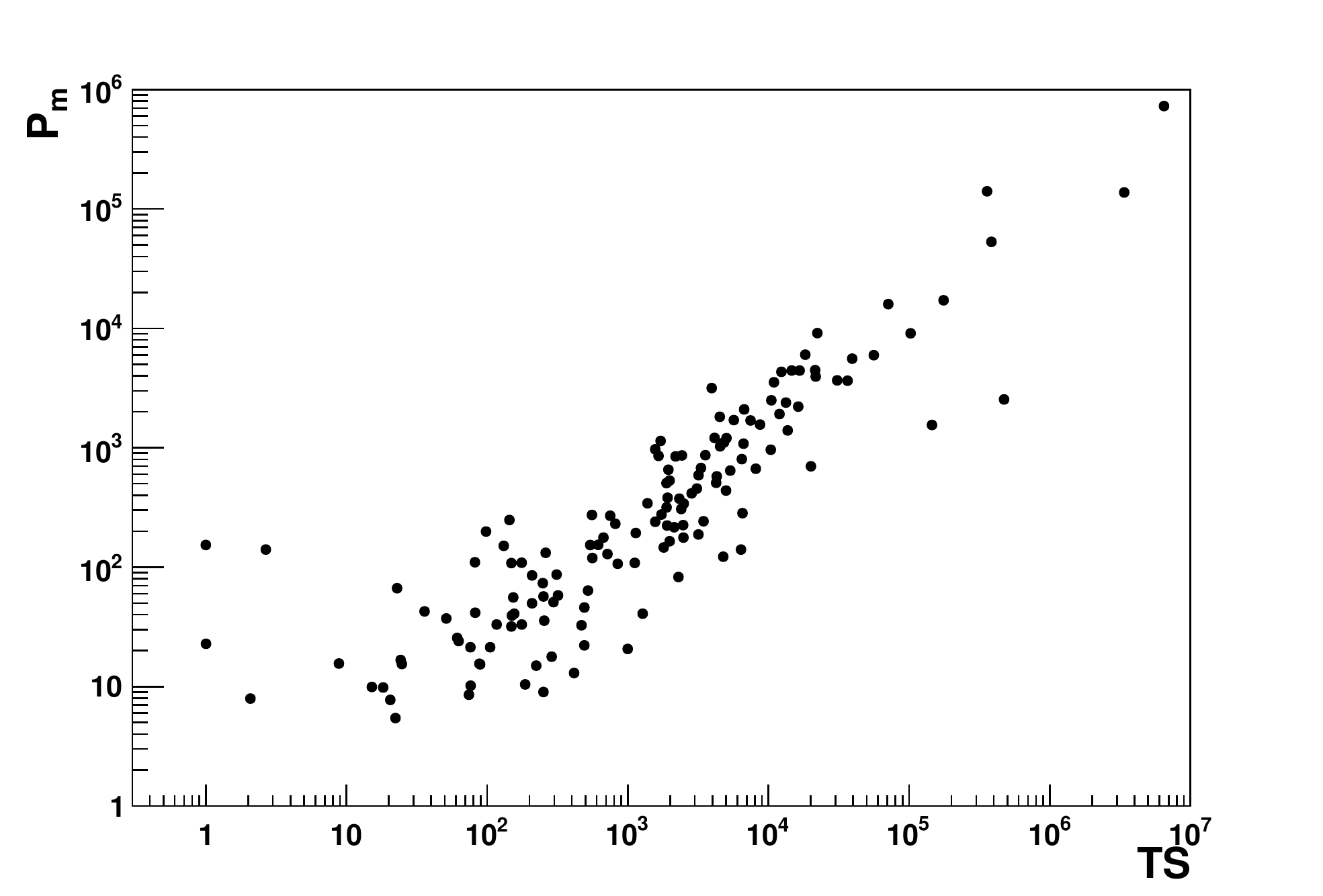}
  \caption{The model weight pulsation significance $P_m$ as a function of TS for the 144 pulsar sample. Pulsation is detected for all pulsars, including the pulsars with TS<25, whose results are reported in Table~\ref{tab:results_faintest}.}
  \label{fig:model_TS}
\end{figure}

The pulsation significance $P_m$ obtained with the model weights is shown in Figure~\ref{fig:model_TS} as a function of TS. All 144 pulsars have $P_m>5.4$, corresponding to $4.6\sigma$. The general trend is that $P_{m}$ on average increases with TS. This is not surprising since, on average, the larger the TS, the stronger the pulsation signal. The scatter around this trend is mainly due to the pulse shape diversity. The most interesting result is the clear detection of pulsation from all pulsars with $\mathrm{TS} \lesssim 50$, which proves that the model weight method is able to detect pulsation even when the pulsar spectrum information is not available or not fully reliable. The $P_m$ and $P_s$ values are given in Table~\ref{tab:results_faintest} for the 12 pulsars with $\mathrm{TS}<25$.

Ignoring the PSF event type information when computing the model weights leads to a loss of sensitivity that decreases with TS, from 20\% on average for faint pulsars to 10\% for the brightest.

\begin{table}
  \centering
  \caption[]{Results of the $\mathrm{TS}<25$ pulsars: TS of the spectral fit, model and simple weight pulsation significances, $P_m$ and $P_s$.}
  \label{tab:results_faintest}

\begin{tabular}{lcccc}
\hline
\noalign{\smallskip}
Pulsar & $(l,b)$ & TS & $P_m$ & $P_s$ \\ 
\noalign{\smallskip}
\hline
\noalign{\smallskip}
J0922+0638 & (225.4,36.4) & 15.2 & 9.9 & 8.4 \\
J1224$-$6407 & (300.0,-1.4) & 24.3 & 16.7 & 13.8 \\
J1455$-$3330 & (330.7,22.6) & 24.8 & 15.4 & 12.7 \\
J1513$-$5908 & (320.3,-1.2) & 2.7 & 140.5 & 144.0 \\
J1646$-$4346 & (341.1,1.0) & 1.0 & 22.8 & 15.0 \\
J1739$-$3023 & (358.1,0.3) & 2.1 & 7.9 & 6.1 \\
J1801$-$2451 & (5.3,-0.9) & 1.0 & 153.4 & 125.4 \\
J1831$-$0952 & (21.9,-0.1) & 8.8 & 15.6 & 10.7 \\
J1832$-$0836 & (23.1,0.3) & 20.5 & 7.7 & 8.0 \\
J1856+0113 & (34.6,-0.5) & 22.9 & 66.7 & 61.3 \\
J1909$-$3744 & (359.7,-19.6) & 18.2 & 9.8 & 6.8 \\
J2317+1439 & (91.4,-42.4) & 22.3 & 5.4 & 4.5 \\
\noalign{\smallskip}
\hline
\end{tabular}
  
\end{table}

\begin{figure}[ht]
  \centering
  \includegraphics[width=9.5cm]{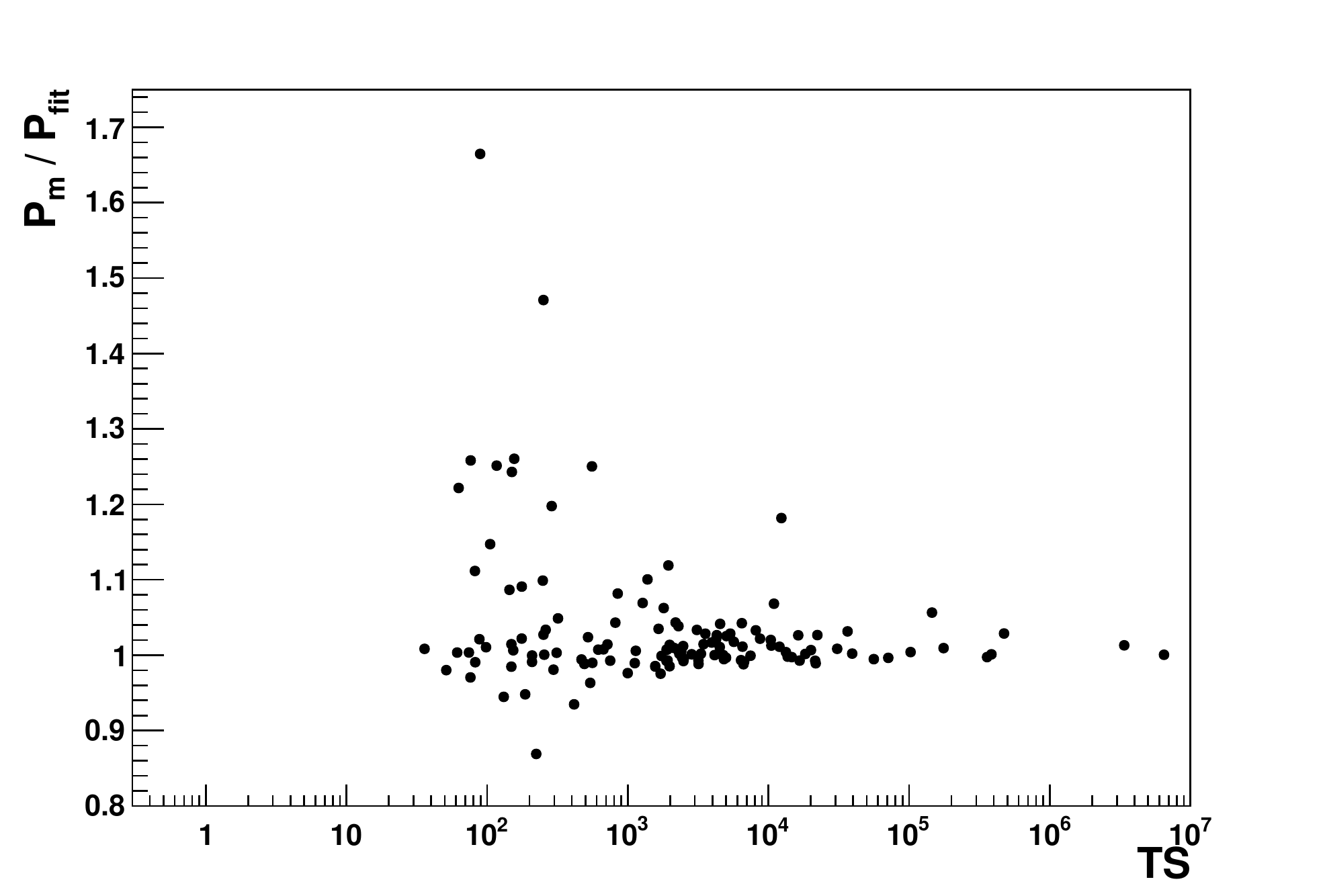}
  \caption{Comparison of the model weight pulsation significance, $P_m$, and the pulsation significance derived when using the pulsar spectrum given by the spectral fit, $P_\mathrm{fit}$, as a function of TS for the 131 pulsars with $\mathrm{TS}>25$ out of the 144 pulsar sample. The two methods give on average the same result but the model weight method is more sensitive by at least 20\% for 8 pulsars, whose results are reported in Table~\ref{tab:results_softandflat}.}
  \label{fig:model_gtlike}
\end{figure}

The comparison between $P_m$ and $P_\mathrm{fit}$ is shown in Figure~\ref{fig:model_gtlike}. On average the model weight method provides the same pulsation significance, within $~\sim$5\% for most of the pulsars. Only one has $P_m$ lower than $P_\mathrm{fit}$ by more than 10\%.

On the contrary, the model weight method often improves significantly over $P_\mathrm{fit}$, with a gain greater than 20\% for 8 pulsars. These 8 pulsars all have $\mathrm{TS}>60$ so it is unlikely that the improvement is explained by a poor estimation of the spectral parameters by the fit. A possible explanation is the presence of a significant off-pulse emission: the spectral parameters derived by the fit correspond to the spectrum of the sum of the pulsed and unpulsed gamma-ray components and not to the pulsed component alone, while the model weight method is able to find spectral parameters closer to the pulsed component spectrum.

As reported in~\cite{2PC}, 34 2PC pulsars have a significant off-peak emission, whose spectrum is compatible with a simple power-law spectrum for 13 of them. Moreover, the off-peak emission of these 13 pulsars is generally soft, with a spectral index $\sim 2$. A flat and soft off-peak emission that is not negligible compared to the pulsed emission could lead to a total emission spectrum significantly different from the pulsed emission spectrum. Investigating that it is also the case of the largest $P_m/P_\mathrm{fit}$ pulsars would require performing the analysis of the off-pulse emission, which is out of the scope of this paper and will be presented in the forthcoming third LAT Gamma-ray Pulsar Catalog (3PC).

%In the case of an unpulsed component with a soft power-law spectrum, the larger the unpulsed component, the softer and flatter the total pulsed+unpulsed spectrum.

 We instead estimate the significance of the curvature of the spectrum. As in~\cite{2PC} and~\cite{3FGL} we perform a maximum-likelihood fit assigning the pulsar a power-law spectrum and compute $\sigma_\mathrm{curv} = (\mathrm{TS}-\mathrm{TS}_\mathrm{PL})^{-1/2}$, where TS and $\mathrm{TS}_\mathrm{PL}$ correspond to the maximum-likelihood fit of a power-law with and without exponential cutoff, respectively. We note however that $\sigma_\mathrm{curv}$ measures the significance of the curvature and not how flat or soft the spectrum is.

A very simple and crude estimator of the softness and flatness of the spectrum is provided by the sum $S = \gamma+\log_{10}E_c$, with $E_c = a^{-1/\beta}$. This estimator can be improved by taking into account the correlation between $\gamma$ and $\log_{10}E_c$. The analysis of the distribution of $\gamma$ vs $\log_{10}E_c$ for the 144 pulsar sample yields a positive correlation with a slope $\sim 1.418$ and the projection along the first principal axis is $S_{f} = 0.576 \gamma + 0.817 \log_{10}E_c$. In the case of a spectrum without significant curvature, the spectral parameters $\gamma$ and $E_c$ are not precisely measured.  To take into account this case, we replace $\gamma$ and $E_c$ by their 68\% upper limit and define the following simple ``soft--and--flat'' estimator:
\begin{equation} \label{eq:softandflat}
  S_{f} = 0.576 (\gamma+\delta\gamma) + 0.817 \log_{10}(E_c+\delta E_c).
\end{equation}

\begin{figure}[ht]
  \centering
  \includegraphics[width=9.5cm]{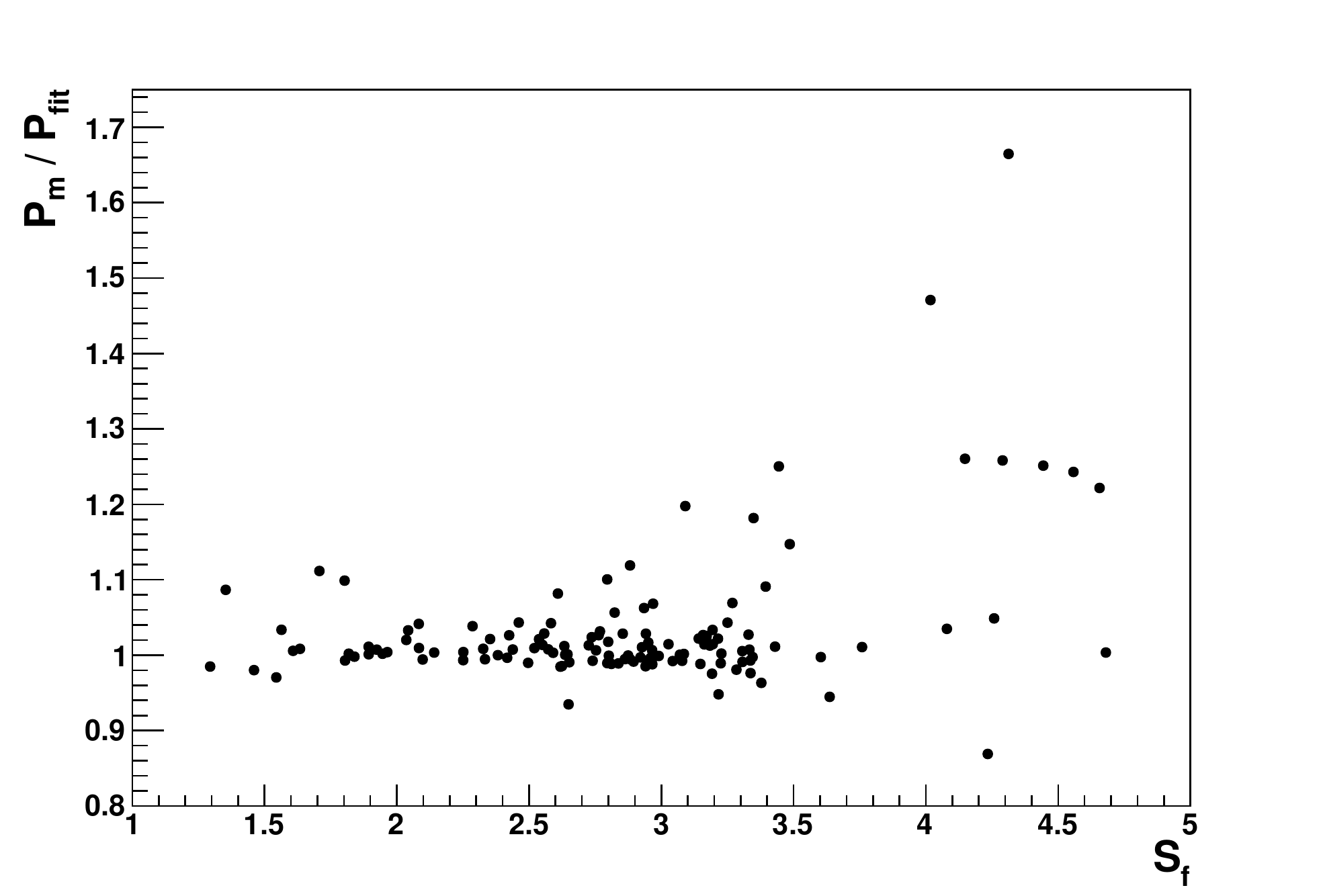}
  \caption{Comparison of the model weight pulsation significance, $P_m$, and the pulsation significance derived when using the pulsar spectrum given by the spectral fit, $P_\mathrm{fit}$, as a function of the ``soft--and--flat'' estimator $S_f$ for the 131 pulsars with $\mathrm{TS}>25$ out of the 144 pulsar sample. The 8 pulsars with $P_m/P_\mathrm{fit}>1.2$ are among the pulsars with the largest $S_f$, {\it i.e.,} with the most soft and flat spectrum.}
  \label{fig:softflatness}
\end{figure}

Figure~\ref{fig:softflatness} shows the ratio $P_m/P_\mathrm{fit}$ as a function of $S_f$. Although the correlation between these two quantities is not perfect, we note that all the 8 pulsars with $P_m/P_\mathrm{fit}>1.2$ lie in the right-hand tail of the $S_f$ distributions, meaning that their spectra are among the most soft and flat of the sample. Some of their properties are given in Table~\ref{tab:results_softandflat}. Only two of them have a curvature significance above $4\sigma$: J0742$-$2822 and J1828$-$1101 with $\sigma_\mathrm{curv} = 8$ and~4.9, respectively. All the others have $\gamma>2.4$. We note that only four of these 8 pulsars are in 2PC and only two are detected as a gamma-ray source (J0742$-$2822 and J1410$-$6132). Out of these two, only J1410$-$6132 is reported to have a significant off-peak emission.

\begin{table}
  \centering
  \caption[]{Pulsars with $\mathrm{TS}>25$ and $P_m/P_\mathrm{fit}>1.2$. The reported spectral index corresponds to the power-law fit, expect for the two $\sigma_\mathrm{curv}>4$ pulsars (indicated with a \textdagger), for which the spectral index is the result of the fit with the power law with an exponential cutoff.}
  \label{tab:results_softandflat}

\begin{tabular}{lccccc}
\hline
\noalign{\smallskip}
Pulsar & TS & $P_m$ & $P_\mathrm{fit}$ & $\gamma$ & $S_f$ \\ 
\noalign{\smallskip}
\hline
\noalign{\smallskip}
J0729$-$1448 & 62.8 & 24.1 & 19.7 & $2.5 \pm 0.1$ & 4.7 \\
J0742$-$2822$^\dagger$ & 556.2 & 273.8 & 219.0 & $1.7 \pm 0.2$ & 3.4 \\
J1151$-$6108 & 150.0 & 39.3 & 31.6 & $2.4 \pm 0.1$ & 4.6 \\
J1410$-$6132 & 116.9 & 33.1 & 26.5 & $2.7 \pm 0.1$ & 4.4 \\
J1431$-$4715 & 76.4 & 10.2 & 8.1 & $2.7 \pm 0.1$ & 4.3 \\
J1531$-$5610 & 155.9 & 40.8 & 32.3 & $2.6 \pm 0.1$ & 4.1 \\
J1828$-$1101$^\dagger$ & 251.4 & 9.0 & 6.1 & $2.1 \pm 0.2$ & 4.0 \\
J1935+2025 & 89.2 & 15.4 & 9.2 & $2.7 \pm 0.1$ & 4.3 \\
\noalign{\smallskip}
\hline
\end{tabular}

\end{table}

\begin{figure}[ht]
  \centering
  \includegraphics[width=9.5cm]{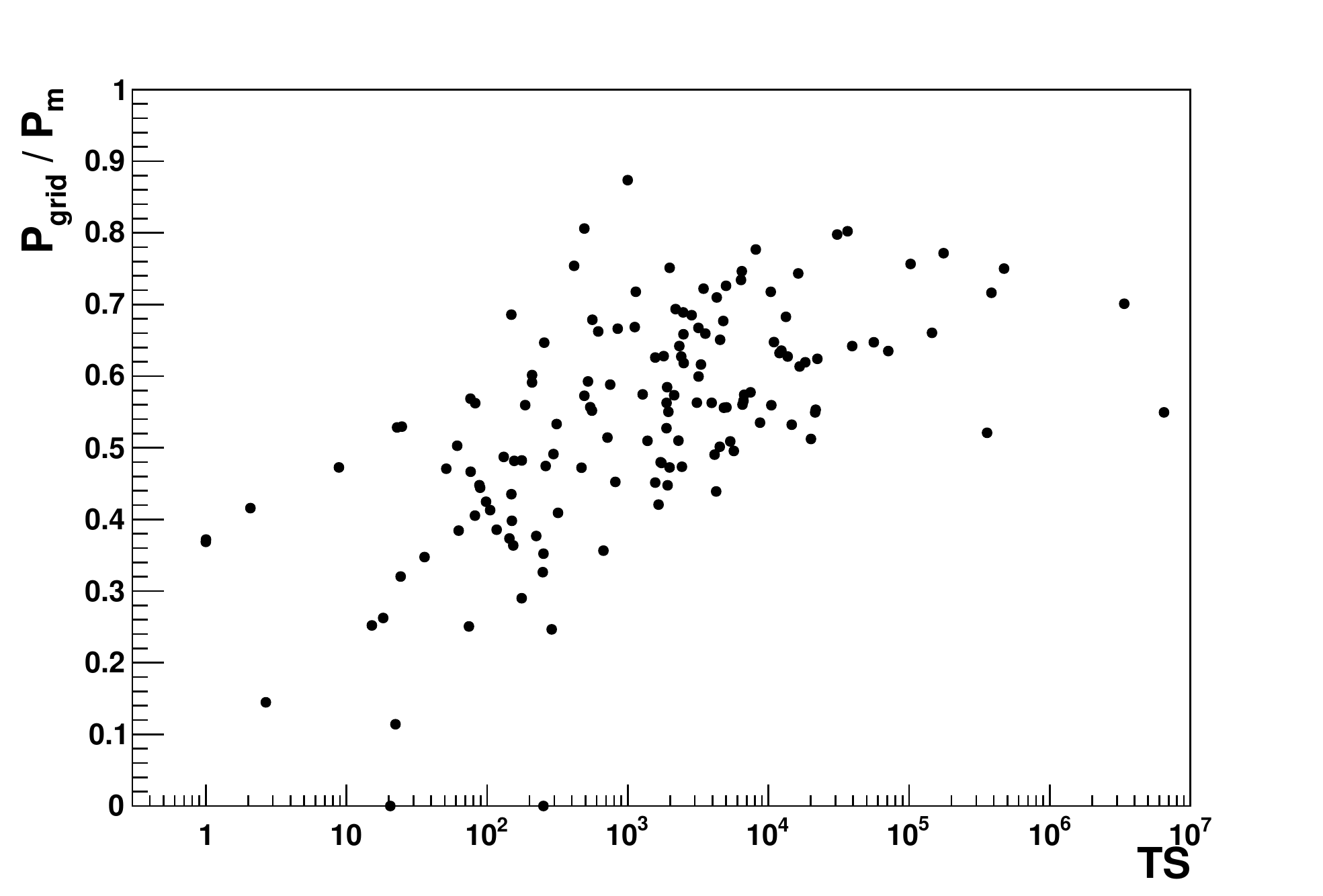}
  \caption{Comparison of the pulsation significance obtained with an unweighted approach, $P_\mathrm{grid}$, and the model weight pulsation significance, $P_m$, as a function of TS for the 144 pulsar sample. The improvement of the model weight method over the unweighted approach is large, especially for the lowest TS pulsars.}
  \label{fig:gridvsmodel}
\end{figure}

The comparison of the model weights with the grid search is shown in Figure~\ref{fig:gridvsmodel} as a function of TS for the 144 pulsar sample. We confirm the improvement of the weighting method over the unweighted approach, with a gain in sensitivity larger than 2 for the $\mathrm{TS}<25$ pulsars.

\subsection{Simple weights}

\begin{figure}[ht]
  \centering
  \includegraphics[width=9.5cm]{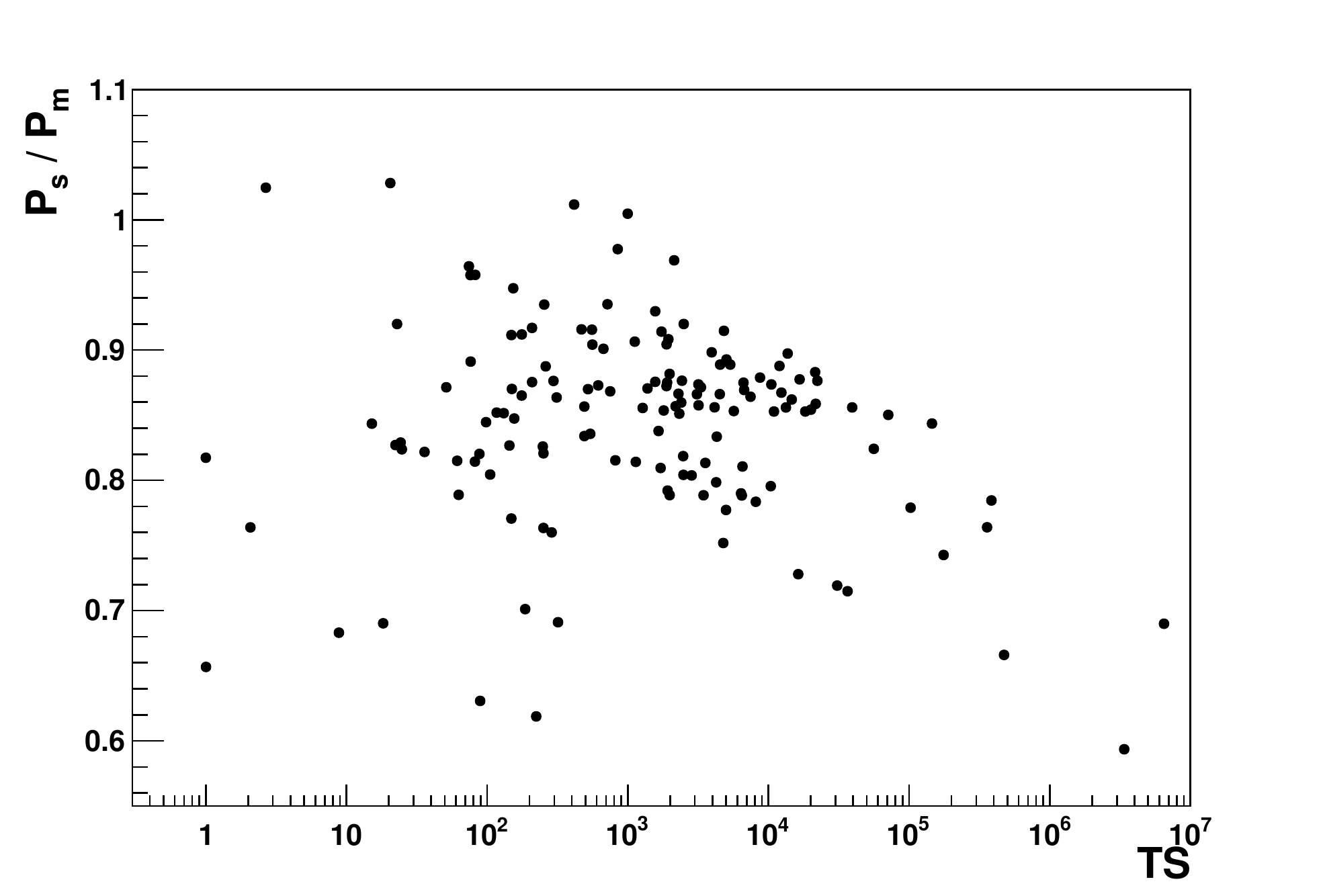}
  \caption{Comparison of the simple and model weight pulsation significances as a function of TS for the 144 pulsar sample. The simple weight method is on average $\sim$15\% less sensitive than the model weight method.}
  \label{fig:simplevsmodel_TS}
\end{figure}

The comparison between the simple weights and the model weights is shown in Figure~\ref{fig:simplevsmodel_TS}. As expected, the simple weights are less powerful than the model weights. The difference in performance decreases from about 30\% for the brightest pulsars to an average of about 15\% at $\mathrm{TS} \sim 300$. This difference never goes beyond 40\%. This difference is relatively small given the simplicity of the simple weight implementation compared to the complex procedure of the model weights. The $P_s$ values are given in Table~\ref{tab:results_faintest} for the 12 pulsars with $\mathrm{TS}<25$. Except for a few of the brightest pulsars, the simple model performs better than the standard grid search, especially for $\mathrm{TS} \leq 100$ where $P_s$ is greater than $P_\mathrm{grid}$ by at least 40\%.

\begin{figure}[ht]
  \centering
  \includegraphics[width=9.5cm]{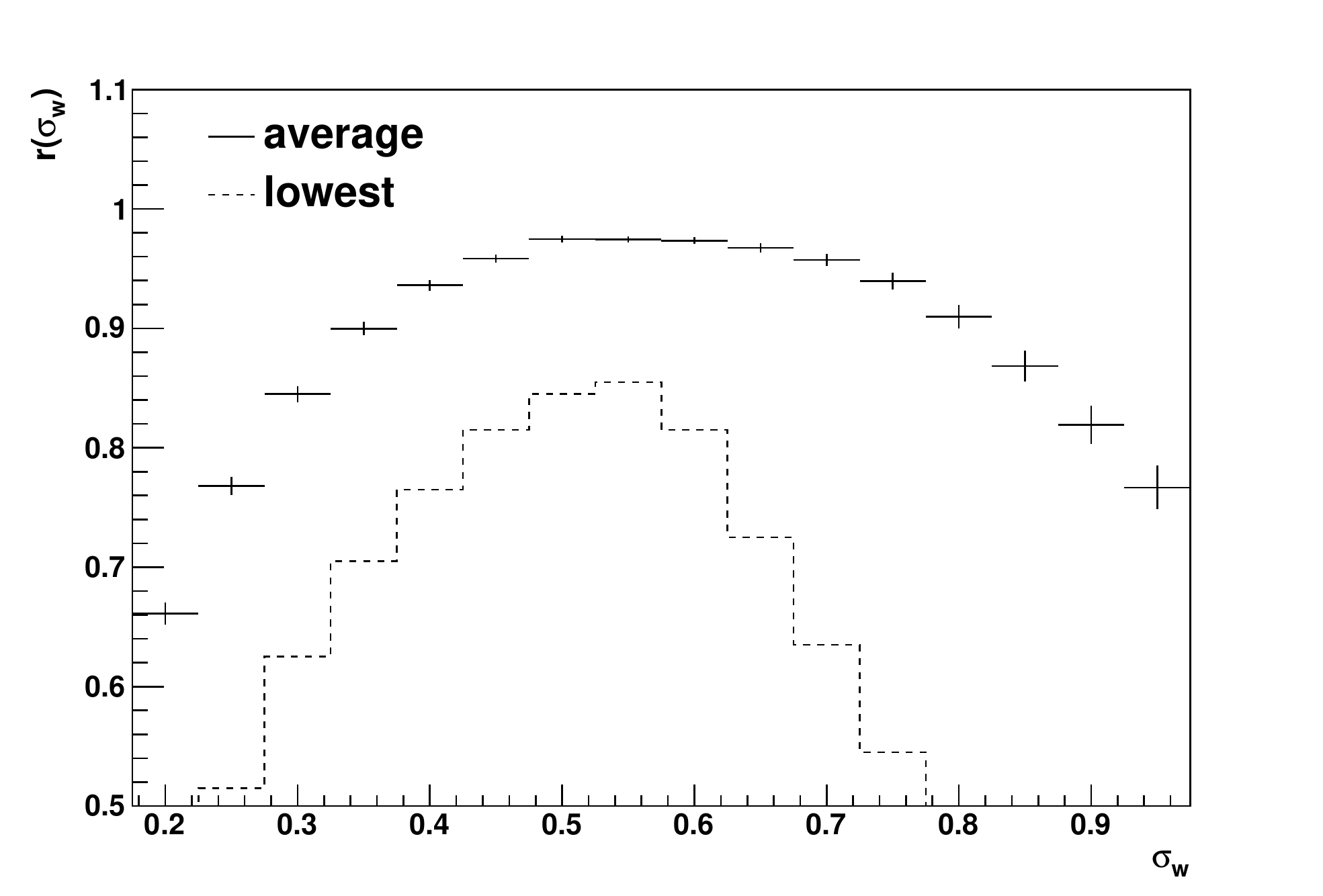}
  \caption{The average (solid) and lowest (dashed) ratio $r(\sigma_w)$ over the 144 pulsar sample as a function of $\sigma_w$. For each $\sigma_w$, $r(\sigma_w)$ is the ratio of $P_s$ divided by the maximum $P_s$ reached over the $\sigma_w$ scan.}
  \label{fig:simple_checksigw}
\end{figure}

When defining the simple weights in Section~\ref{sec:simple_weightdef}, we set $\sigma_w = 0.5$. To validate this choice, we perform a scan over $\sigma_w$ between 0.2 and 0.95, with a 0.05 step. For each pulsar and each value of $\sigma_w$, we compute $r(\sigma_w)$, the ratio of $P_s$ divided by the maximum $P_s$ reached over the $\sigma_w$ scan. The average over the 144 pulsar sample of this ratio as a function of $\sigma_w$ is shown in Figure~\ref{fig:simple_checksigw}, as well as the lowest ratio, {\it i.e.} corresponding to the pulsar for which the choice of $\sigma_w$ is the worst. The two quantities have a maximum between 0.5 and 0.6 and do not vary much around the maximum, which validates the $\sigma_w = 0.5$ choice.

\begin{figure}[ht]
  \centering
  \includegraphics[width=9.5cm]{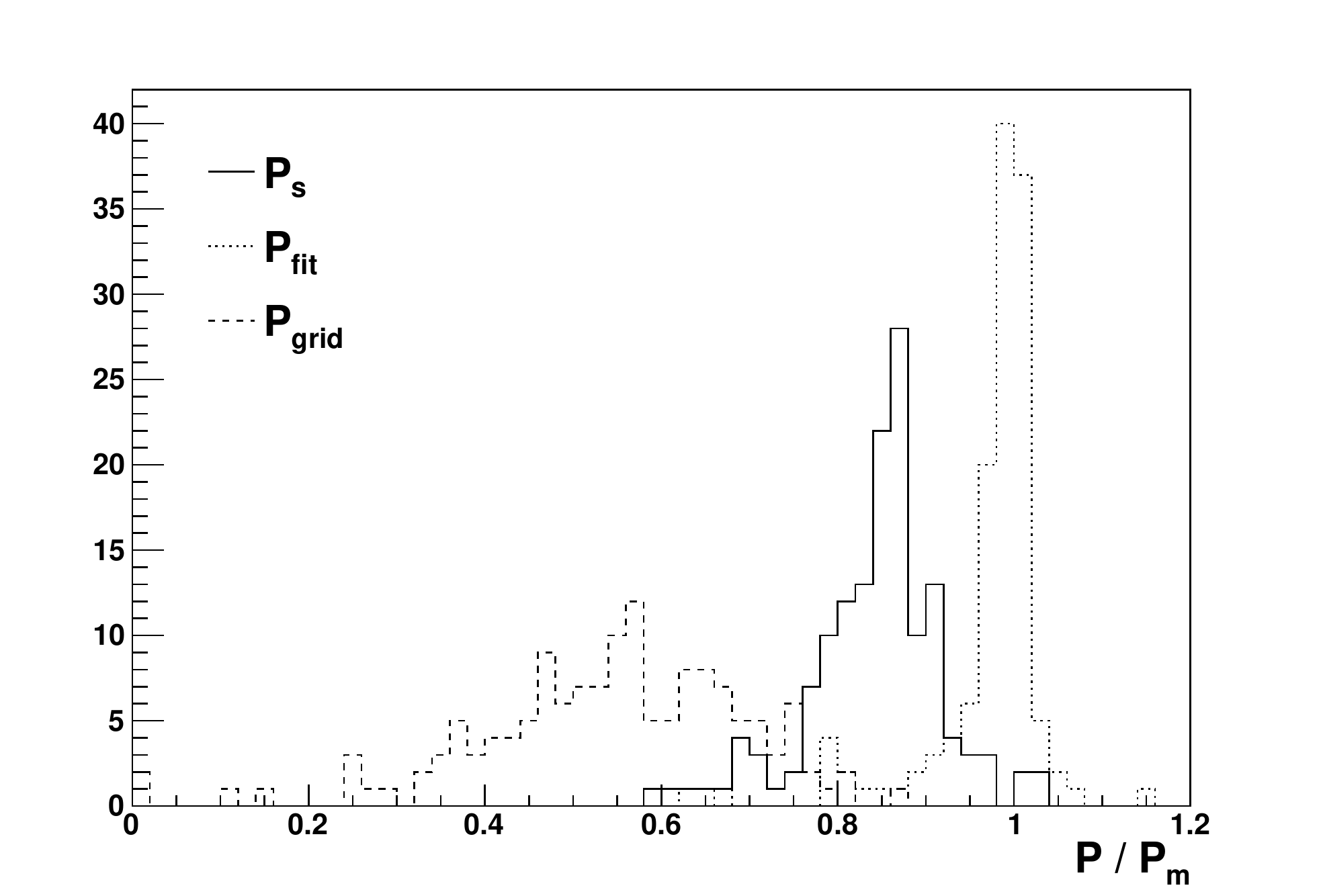}
  \caption{Ratio of the pulsation significance with respect to $P_m$ for the simple weights (solid), original event weight (dotted) and the grid search (dashed), for the 144 pulsar sample.}
  \label{fig:allcomp}
\end{figure}

To summarize the results, we show in Figure~\ref{fig:allcomp} the comparison of all methods that allows a clear ranking of the pulsation search methods, from the least sensitive unweighted grid scan to the most powerful \cite{kerr2011} and model weight methods, with the latter, contrary to the former, being able to detect pulsation from very faint pulsars.

We note that both the simple and model weight approaches can be used at other wavelengths ({\it e.g.} X-ray and TeV bands) but they need to be adapted to the specific context of each wavelength, taking into account the pulsar spectral parameter phase space, the PSF energy dependence and the typical background spectrum. In the case of the simple weight method, the derived general shape of $f(E)$ may be very different from the gaussian-like one obtained for the LAT energy band.

\section{Conclusion} \label{sec:conclusion}

We have shown that it is possible to extend the event weighting technique to very faint gamma-ray sources when searching for pulsation and presented two approaches. The first one, the simple weight method, uses a very simple definition of the weights while the second one, the model weight method, fully takes into account the spatial and spectral information of the RoI around the pulsar. The key point for both methods is to explore efficiently the pulsar spectral parameter space, which is done with only 6 trials.

The model weight method reaches the same perfomance as the original event weighting method that was not applicable to the very faint gamma-ray sources. It can even be more powerful in the case of pulsars with a significant off-pulse emission.

The simple weight method is less sensitive than the model weight one but the loss of sensitivity is only $\sim$30\% for faint sources. So this simple approach can be very useful, especially since it is straightforward to implement and is much less CPU intensive.

As a consequence the model weights method is on average the most sensitive method. It works for all pulsars (faint or bright, with or without off-pulse emission) and naturally benefits from any improvement of the instrument performance ({\it e.g.} Pass~8 PSF event types).

After ten years in orbit, {\it Fermi}-LAT continues to take data that are folded with the updated ephemerides provided by the Pulsar Timing Consortium. The two new methods presented in this paper are designed to help to detect new gamma-ray pulsars, including very faint ones, allowing us, for instance, to further test the existence of a spin-down power ``deathline''~\citep{smith2018} below which pulsars might cease to produce gamma-ray emission.

%%For a given source, the gain in pulsation sensitivity goes as square root of time. This is true for both methods but in the case of the model weights, the gain also comes from a better description of the sources around the pulsar. This might allow us to do slightly better than the square root of time limit.

\begin{acknowledgements}
We thank our {\it Fermi}-LAT collaborators Toby Burnett, Lucas Guillemot, David Smith and Matthew Kerr for fruitful discussions.

The Nan\c cay Radio Observatory is operated by the Paris Observatory, associated with the French Centre National de la Recherche Scientifique (CNRS).

The Parkes radio telescope is part of the Australia Telescope 533 which is funded by the Commonwealth Government for operation as a National Facility managed by CSIRO. We thank our colleagues for their assistance with the radio timing observations.

The Lovell Telescope is owned and operated by the University of Manchester as part of the Jodrell Bank Centre for Astrophysics with support from the Science and Technology Facilities Council of the United Kingdom.

The \textit{Fermi} LAT Collaboration acknowledges generous ongoing support
from a number of agencies and institutes that have supported both the
development and the operation of the LAT as well as scientific data analysis.
These include the National Aeronautics and Space Administration and the
Department of Energy in the United States, the Commissariat \`a l'Energie Atomique
and the Centre National de la Recherche Scientifique / Institut National de Physique
Nucl\'eaire et de Physique des Particules in France, the Agenzia Spaziale Italiana
and the Istituto Nazionale di Fisica Nucleare in Italy, the Ministry of Education,
Culture, Sports, Science and Technology (MEXT), High Energy Accelerator Research
Organization (KEK) and Japan Aerospace Exploration Agency (JAXA) in Japan, and
the K.~A.~Wallenberg Foundation, the Swedish Research Council and the
Swedish National Space Board in Sweden.
 
Additional support for science analysis during the operations phase is gratefully
acknowledged from the Istituto Nazionale di Astrofisica in Italy and the Centre
National d'\'Etudes Spatiales in France. This work performed in part under DOE
Contract DE-AC02-76SF00515.
\end{acknowledgements}

\begin{appendix}

\section{Monte Carlo estimated probability distribution of $H_{20}$} \label{app:htestcalib}

The asymptotic $H_{20}$ probability distribution has been estimated using Monte Carlo (MC) by~\cite{dejager1989} and~\cite{dejager2010}. It corresponds to a simple exponential $P(H_{20}>x) = e^{-\lambda x}$ with $\lambda \sim 0.4$. This result is very close to the analytic solution found by~\cite{kerr2011} that yields a practical formula (valid for $m \geq 10$) with $\lambda = 0.398405$.

\begin{figure}[ht]
  \centering
  \includegraphics[width=9.5cm]{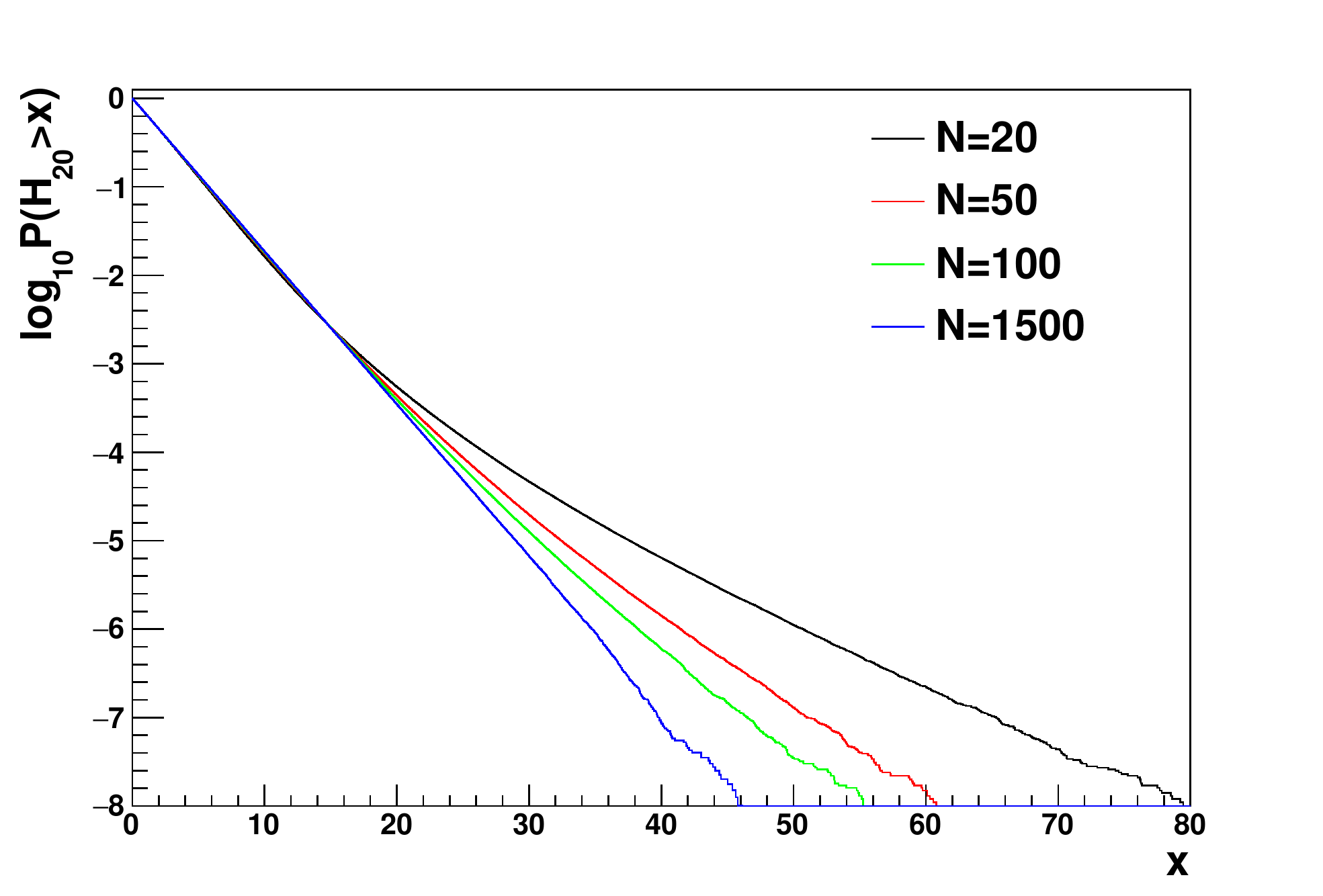}
  \caption{The logarithm of the $H_{20}$ cumulative distribution for various data sample sizes.}
  \label{fig:htestcalib0}
\end{figure}

Pulsation searches very often require multiple trials corresponding to different event selections. Some of these can lead to small data samples. As a consequence it is important to estimate the $H_{20}$ probability distribution for low $N$, the number of events used when computing $H_{20}$. To do so, we ran $10^9$ MC realizations for various $N$, assigning a random phase to each event. Figure~\ref{fig:htestcalib0} shows the logarithm of the cumulative distribution, $\log_{10}P(H_{20}>x)$, for $N=20,50,100$ and 1500. The asymptotic behavior is reached in $N=1500$ case but the distribution of the other cases shows a clear departure from the pure exponential above $x \sim 20$.

\begin{figure}[ht]
  \centering
  \includegraphics[width=9.5cm]{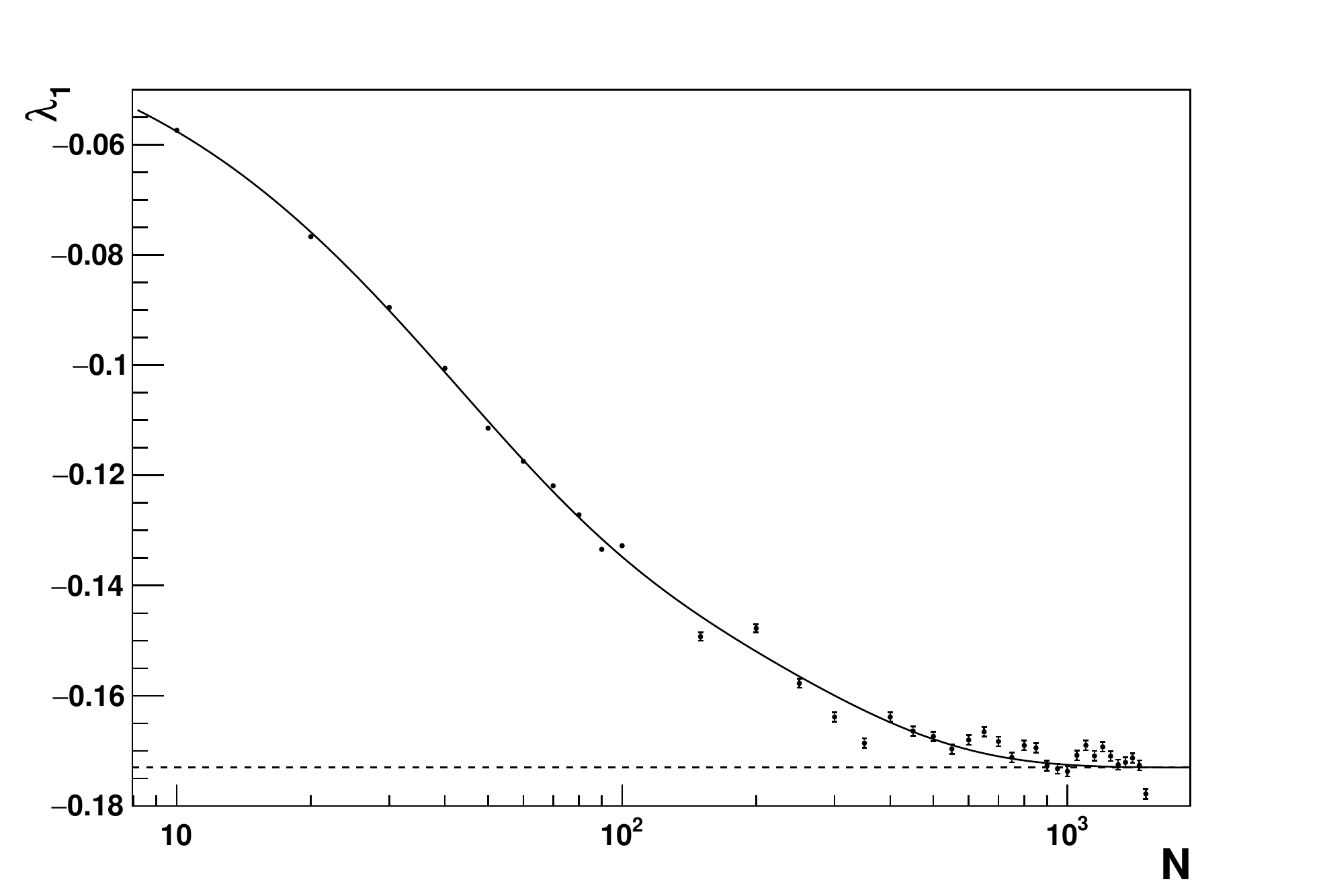}
  \caption{The parameterization of $\lambda_1$, the slope of $\log_{10}P(H_{20}>x)$ over the interval corresponding to $-7<\log_{10}P(H_{20}>x)<-4$, as a function of $N$.}
  \label{fig:htestcalib1}
\end{figure}

To characterize this departure, we fit $\log_{10}P(H_{20}>x)$ with a linear polynomial over the interval corresponding to $-7<\log_{10}P(H_{20}>x)<-4$ ({\it i.e.} $x \gtrsim 23$). The slope $\lambda_{1}$ of this linear polynomial is shown in Figure~\ref{fig:htestcalib1} as a function of $N$. We find the following parameterization for $\lambda_1$ (shown in Figure~\ref{fig:htestcalib1}):

\begin{equation} \label{eq:lambda1}
\lambda_1(N) = \lambda_0 + 0.0525796 e^{-N/215.170} + 0.086406 e^{-N/35.5709} 
\end{equation}
where $\lambda_0 = -0.398405/\log(10) = -0.173025$.

\begin{figure}[ht]
  \centering
  \includegraphics[width=9.5cm]{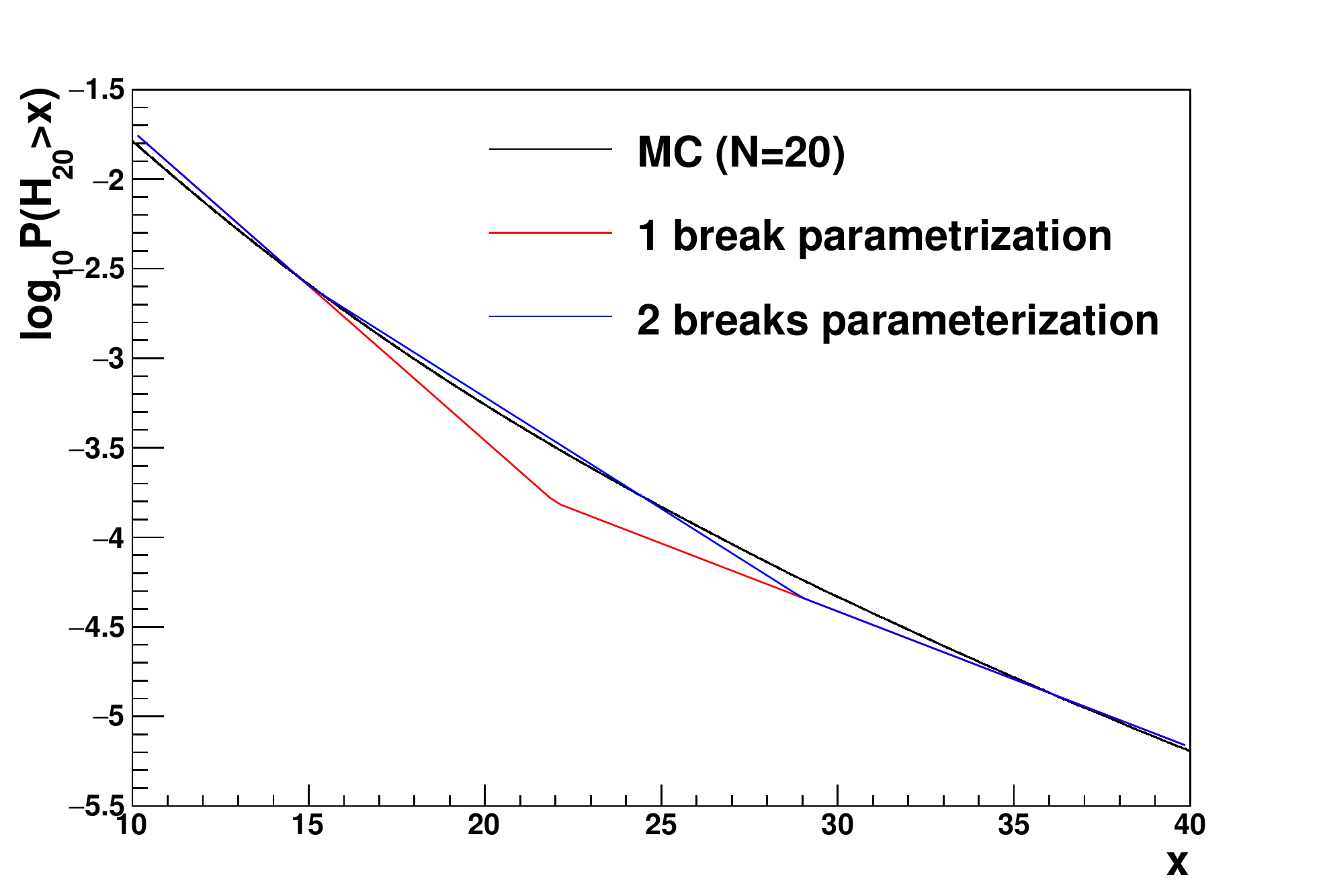}
  \caption{The logarithm of the $H_{20}$ cumulative distribution for $N=20$ in the $x$ range where a single broken linear polynomial does not provide a good approximation. The black line corresponds to the MC result while the red and blue curves correspond to the single and double broken linear polynomial approximations, respectively.}
  \label{fig:htestcalib2}
\end{figure}

We first approximate $\log_{10}P(H_{20}>x)$ with a broken linear polynomial whose slopes at small and large $x$ are $\lambda_0$ and $\lambda_1(N)$, respectively. We find that the break position is compatible with $x=22$ for all $N$. For low values of $N$ the broken linear polynomial approximation does not work well around the break position, as shown in Figure~\ref{fig:htestcalib2}. To have a better representation of $\log_{10}P(H_{20}>x)$, we use a double broken linear polynomial approximation, with the same slopes at small and large $x$ as for the single one, with break positions at $x=15$ and $x=29$:

\begin{equation} \label{eq:htestcalib}
\log_{10}P(H_{20}>x) \sim 
\left\{
\begin{array}{ll}
      \lambda_0 x & \text{if } x < 15,\\
      15\lambda_0 + \frac{\lambda_0 +\lambda_1(N)}{2} (x-15) & \text{if } 15<x<29\\
      22\lambda_0 + \lambda_1(N) (x-22) & \text{if } x>29.
\end{array}
\right.
\end{equation}

When considering the $x$ interval such that $P(H_{20}>x)>10^{-7}$, the maximum of the absolute difference between the parameterization and the MC result is less than 0.1 for $N \geq 20$. For $N=10$, the maximum difference reaches 0.25 at $x=29$. As a consequence, the approximation given by Equation~\ref{eq:htestcalib} is valid for $N \geq 20$ with a 0.1 precision for $P(H_{20}>x)>10^{-7}$.

\section{Monte Carlo estimated probability distribution of $H_{20w}$} \label{app:whtestcalib}

To estimate the probability distribution of $H_{20w}$, we ran $10^8$ MC realizations of a 5~degree region of interest with a uniform background following the Galactic diffuse emission spectrum defined in Section~\ref{sec:simple_weightdef} (corresponding to a broken power law, with spectral indices of $\sim 1.6$ and $\sim 2.5$ below and above $\sim 3$~GeV, respectively), convoluted with the {\it Fermi}-LAT effective area. These realizations are performed for various numbers of events. For each realization, we compute $H_{20w}$ using the simple weights definition of Equation~\ref{eq:weightdef} with $\sigma_\mathrm{psf}$ set to 1~degree independently of energy. For CPU efficiency's sake, each realization is used twice, with $\mu_w$ set to 2.5 and 3, each choice leading to a value of $H_{20w}$.

\begin{figure}[ht]
  \centering
  \includegraphics[width=9.5cm]{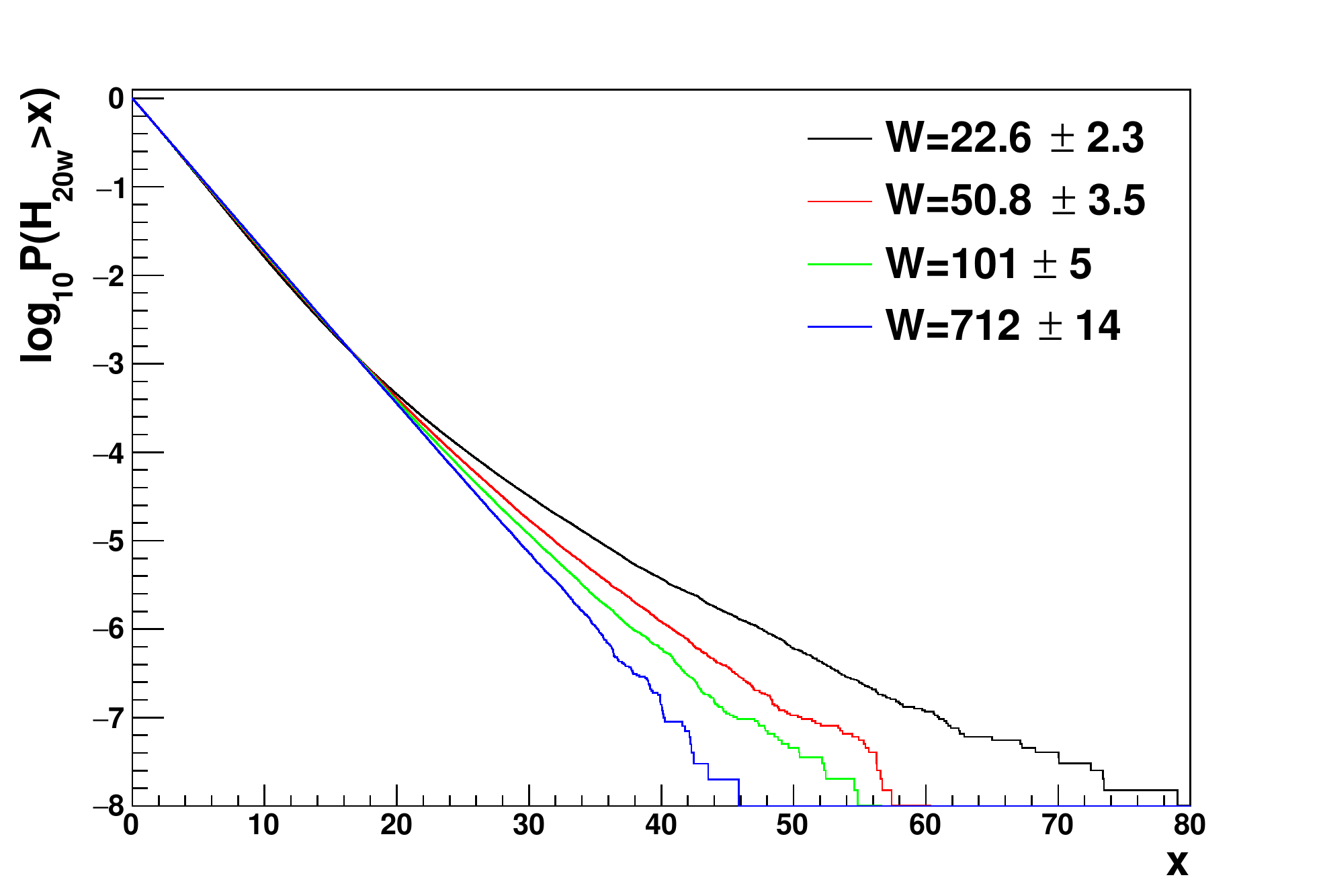}
  \caption{$\log_{10}P(H_{20w}>x)$ for various sum of weights.}
  \label{fig:htestwcalib0}
\end{figure}

When parameterizing $\log_{10}P(H_{20}>x)$ in the previous Section, we use the number of events because it drives the level of fluctuations. In the case of the weighted version of $H_\mathrm{test}$, the number of events is not useful anymore. We use instead the sum of the weights, $W$, that we compute under the prescription that the maximum weight is 1. Figure~\ref{fig:htestwcalib0} shows $\log_{10}P(H_{20w}>x)$ for some of the MC configurations corresponding to $W \sim 20,50,100,700$. As in the unweighted case, $\log_{10}P(H_{20w}>x)$ departs from a pure exponential at low $W$ and can be approximated to first order by a broken linear polynomial. We note that the various MC configurations (number of events, $\mu_w$ choice) were chosen to explore a range for $W$ between about 10 and 1000.

\begin{figure}[ht]
  \centering
  \includegraphics[width=9.5cm]{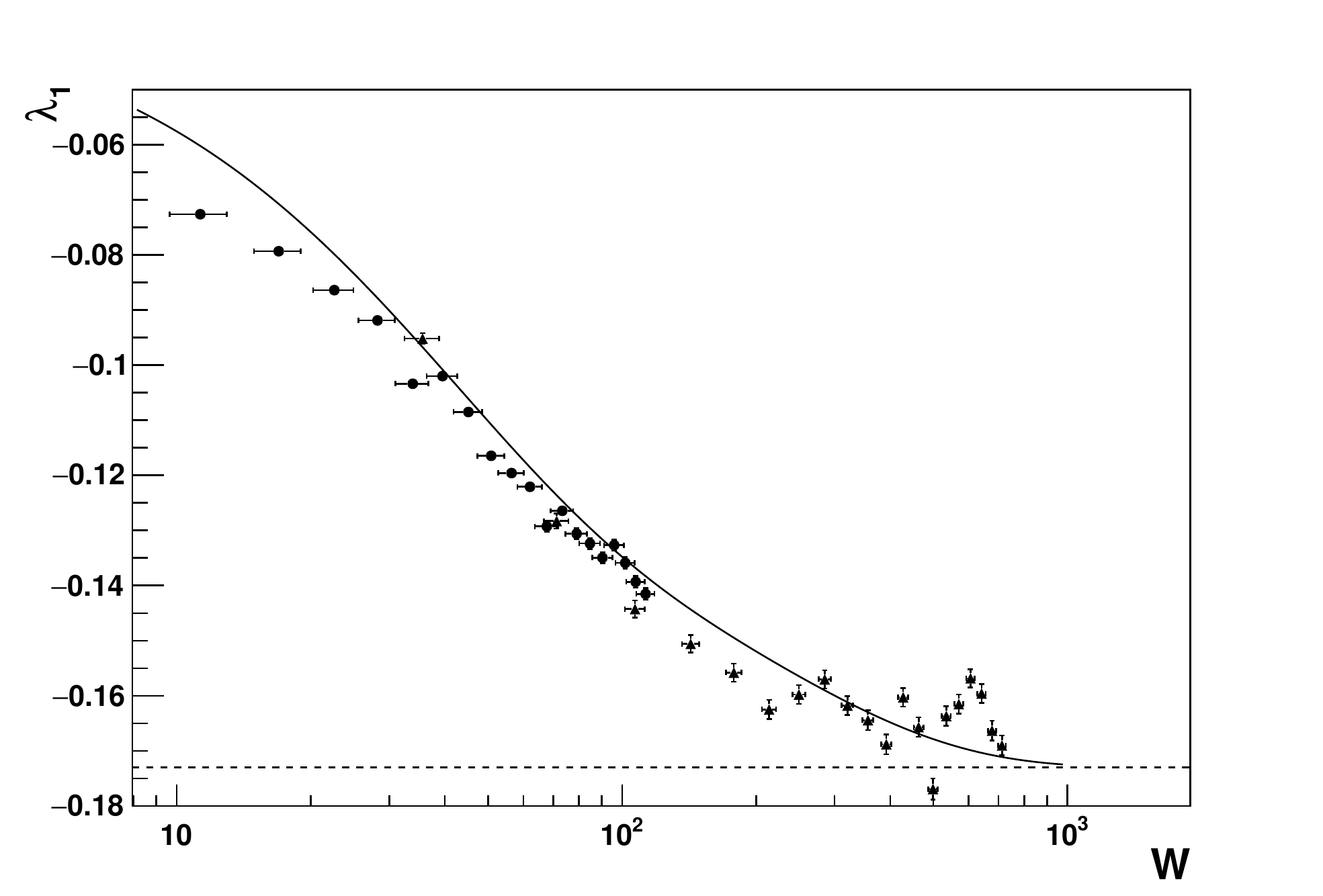}
  \caption{The slope of $\log_{10}P(H_{20w}>x)$ over the interval corresponding to $-7<\log_{10}P(H_{20w}>x)<-4$ as a function of $W$. Circles and triangles correspond to $\mu_w=2.5$ and 3, respectively.}
  \label{fig:htestwcalib1}
\end{figure}

\begin{figure}[ht]
  \centering
  \includegraphics[width=9.5cm]{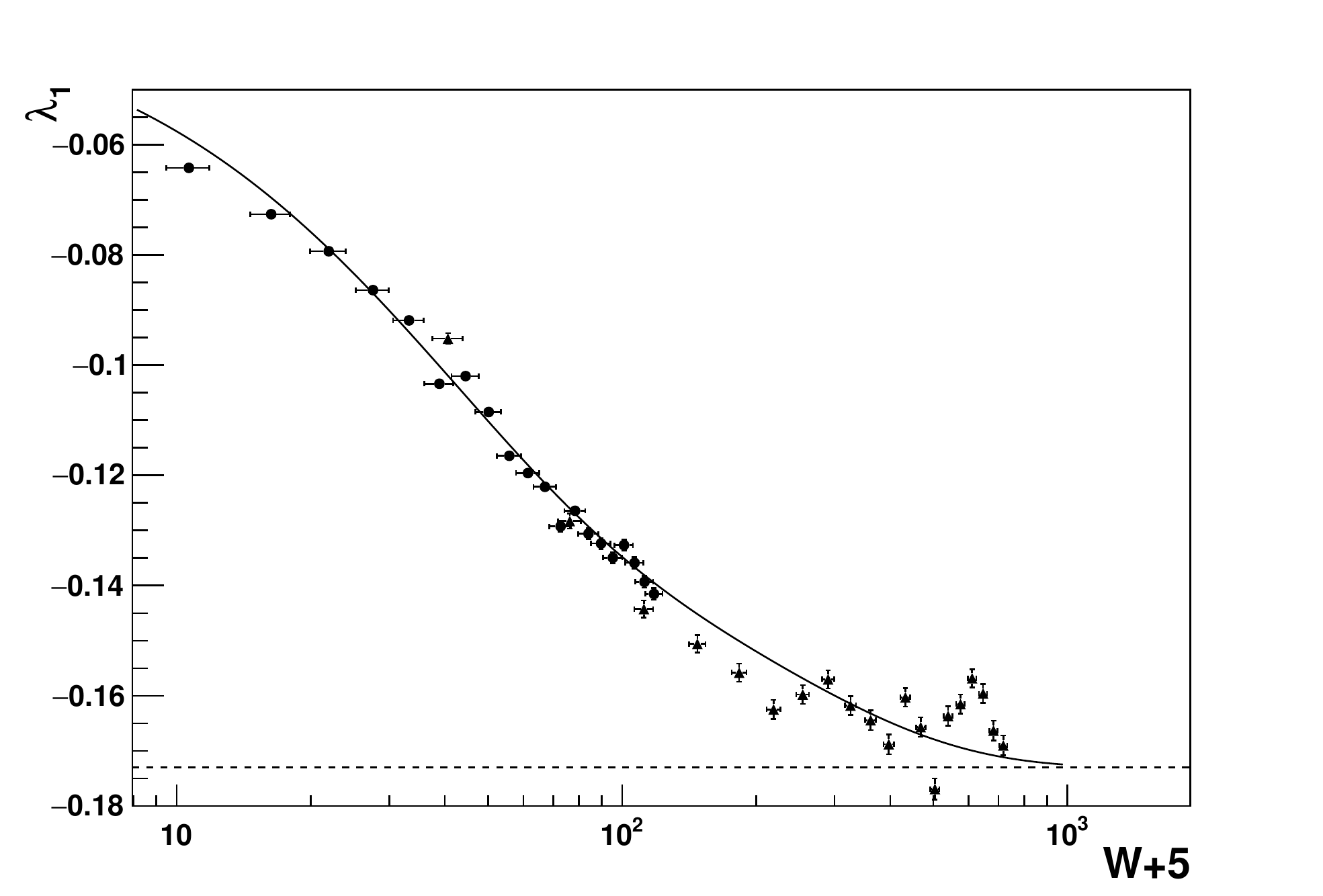}
  \caption{The slope of $\log_{10}P(H_{20w}>x)$ over the interval corresponding to $-7<\log_{10}P(H_{20w}>x)<-4$ as a function of $W+5$.  Circles and triangles correspond to $\mu_w=2.5$ and 3, respectively.}
  \label{fig:htestwcalib2}
\end{figure}

We estimate $\lambda_1$, the slope over the interval corresponding to $-7<\log_{10}P(H_{20w}>x)<-4$. Figure~\ref{fig:htestwcalib1} shows how the $\lambda_1$ variation with $W$ compares to the unweighted-case $\lambda_1$ parameterization of Equation~\ref{eq:lambda1}. This parameterization works well for large values of $W$ but not at low $W$. We find that using $W+5$ instead of $W$ gives a better match, as shown in Figure~\ref{fig:htestwcalib2}. The fact that increasing $W$ leads to a better match is not surprising since, in the weighted case, the number of events that play a significant role in the $H_{20w}$ computation ({\it i.e.} with a relatively large weight) is on average larger than $W$.

As a consequence, we use the following parameterization of $\log_{10}P(H_{20w}>x)$, that is obtained by simply replacing $N$ by $W+5$ in Equation~\ref{eq:htestcalib}:
\begin{equation} \label{eq:htestwcalib}
\log_{10}P(H_{20w}>x) \sim 
\left\{
\begin{array}{ll}
      \lambda_0 x & \text{if } x < 15,\\
      15\lambda_0 + \frac{\lambda_0 +\lambda_1(W+5)}{2} (x-15) & \text{if } 15<x<29\\
      22\lambda_0 + \lambda_1(W+5) (x-22) & \text{if } x>29.
\end{array}
\right.
\end{equation}

When considering the $x$ interval with $P(H_{20w}>x)>10^{-7}$, the maximum of the absolute difference between the parameterization and the MC result is less than 0.1 for $W \geq 10$. As a consequence, the approximation given by Equation~\ref{eq:htestwcalib} is valid for $W \geq 10$ with a 0.1 precision for $P(H_{20w}>x)>10^{-7}$.

\end{appendix}


\begin{thebibliography}{}
\bibitem[Abdo et al.(2013)]{2PC} Abdo, A. A., Ajello, M., Allafort, A., et al. 2013, ApJS, 208, 17
\bibitem[Abdo et al.(2010)]{B1509} Abdo, A. A., Ackermann, M. Ajello, M., et al. 2010, ApJ, 714, 1
\bibitem[Acero et al.(2015)]{3FGL} Acero, F., Ackermann, M., Ajello, M., et al. 2015, ApJS, 218, 23
\bibitem[Ackermann et al.(2013)]{PSFDETERMINATION} Ackermann, M., Ajello, M., Allafort, A., et al. 2013, ApJ, 765, 1
\bibitem[Atwood et al.(2009)]{latinstrument} Atwood, W. B., Abdo, A. A., Ackermann, M., et al. 2009, ApJ, 697, 1071
\bibitem[Atwood et al.(2013)]{pass8} Atwood, W., Albert, A., Baldini, L., et al. 2013, eConf C121028, 8, in Proc. 4th Fermi Symposium, Monterey
\bibitem[Bickel et al.(2008)]{BKR08} Bickel, P., Kleinj, B., Rice, J. 2008, ApJ, 685, 384
\bibitem[Bruel et al.(2018)]{pass8P8R3} Bruel, P., Burnett, T. H., Digel, S. W., et al. 2018, presented at the 8th Fermi symposium, arXiv:1810.11394 [astro-ph.IM]
\bibitem[Clark et al.(2017)]{clarck2017} Clark, C. J., Wu, J., Pletsch, H. J., et al. 2017, ApJ, 834, 106
\bibitem[de Jager et al.(1989)]{dejager1989} de Jager, O. C., Raubenheimer, B. C., Swanepoel, J. W. H. 1989, A\&A, 221, 180
\bibitem[de Jager et al.(2010)]{dejager2010} de Jager, O. C., B\" usching, I. 2010, A\&A, 517, L9
\bibitem[Edwards et al.(2006)]{edwards2006} Edwards, R. T., Hobbs, G. B., Manchester, R. N. 2006, MNRAS, 372, 1549
\bibitem[Hobbs et al.(2006)]{hobbs2006} Hobbs, G. B., Edwards, R. T., Manchester, R. N. 2006, MNRAS, 369, 655
\bibitem[Hou et al.(2014)]{hou2014} Hou, X., Smith, D. A., et al. 2014, A\&A, 570, A44
\bibitem[Kerr(2011)]{kerr2011} Kerr, M. 2011, ApJ, 732, 38
\bibitem[Kuiper et al.(2017)]{kuiper2017} Kuiper. L., Hermsen, W., Dekker, A. 2017, MNRAS, 475, 1
\bibitem[Laffon et al.(2014)]{laffon2014} Laffon, H., Smith, D. A., Guillemot, L. et al. 2014, in Proc. 5th Fermi symposium, Nagoya
\bibitem[Mattox et al.(1996)]{mattox} Mattox, J. R., Bertsch, D. L., Chiang, J., et al. 1996, ApJ, 461, 396
\bibitem[Moffat(1969)]{moffat} Moffat, A. F. J. 1969, A\&A, 3, 455
\bibitem[Pleunis et al(2017)]{pleunis2017} Pleunis, Z., Bassa, C. G., Hessels, J. W. T., et al. 2017, ApJL, 846, L19
\bibitem[Ray et al.(2011)]{ray2006}  Ray, P. S., Kerr, M., Parent, D., et al. 2011, ApJS, 194, 17
\bibitem[Smith et al.(2008)]{PTC} Smith, D. A., Guillemot, L., Camilo, F., et al. 2008, A\&A, 492, 923
\bibitem[Smith et al.(2017)]{smith2017} Smith, D. A., Guillemot, L., Kerr, M., et al. 2017, in Proc. 11th INTEGRAL conference, Amsterdam
\bibitem[Smith et al.(2018)]{smith2018} Smith, D. A., Bruel, P., Cognard, I., et al. 2018, ApJ in press, arXiv:1812.00719 [astro-ph.HE]
\bibitem[Thompson(2008)]{thompson2008} Thompson, D. J. 2008, Reports on Progress in Physics, 71, 116901
\end{thebibliography}
\end{document}